\documentclass[twocolumn]{aastex631}

\usepackage{amssymb}
\usepackage{amsmath}
\usepackage{graphicx}
\usepackage{float}
\usepackage{physics}
\let\tablenum\relax
\usepackage{siunitx}

\DeclareSIUnit{\elecd}{\meter\squared\per\second}

\makeatletter
\newcommand\newtag[2]{#1\def\@currentlabel{#1}\label{#2}}
\makeatother

\begin{document}

\title{Onset of Plasmoid Reconnection during Magnetorotational Instability}  

\author{Jarrett Rosenberg}
\affiliation{Princeton Plasma Physics Laboratory, Princeton University, Princeton, New Jersey 08543, USA}
\affiliation{{Department of Physics, Applied Physics, and Astronomy, Rensselaer Polytechnic Institute, Troy NY, 12180, USA}}

\author{Fatima Ebrahimi}
\affiliation{Princeton Plasma Physics Laboratory, Princeton University, Princeton, New Jersey 08543, USA}
\affiliation{Department of Astrophysical Sciences, Princeton University, Princeton, New Jersey 08544, USA}


\begin{abstract}
The evolution of current sheets in accretion flows undergoing magnetorotational instability (MRI) is examined through two and three dimensional numerical modelling of the resistive MHD equations in global cylindrical geometry. With an initial uniform magnetic field aligned in the vertical ($z$) direction, MRI produces radially extended toroidal (azimuthal) current sheets. In both 2D and 3D when axisymmetric modes dominate, these current sheets attract each other and merge in the poloidal ($rz$) plane, driving magnetic reconnection when the Lundquist number $S > \num{3e2}$, making it a possible source of plasmoids (closed magnetic loops) in accretion disks. At high Lundquist numbers in the 2D regime, starting at $S = \num{5e3}$, self-consistent MRI-generated current sheets become thin and subject to plasmoid instability, and therefore spontaneous magnetic reconnection. When non-axisymmetric 3D modes dominate, turbulence makes the azimuthal current sheets further unstable, and stretch vertically. Toroidally extended vertical current sheets in the inner region, as well as larger 3D magnetic islands in the outer regions of the disks are also formed. 
{These findings have strong ramifications for astrophysical disks as potential sources of plasmoids that could cause local heating, particle acceleration, and high energy EM radiation.}
\end{abstract}

\section{Introduction}

Accretion disks power many of the most luminous astrophysical objects, including X-ray binaries, accreting black hole sources, quasars, and gamma-ray bursts. As matter accretes from astrophysical disks onto a central body, angular momentum is rapidly transported outward. In these differentially rotating systems, weak magnetic fields can trigger turbulence via the magnetorotational instability (MRI), transporting angular momentum outward by harnessing Maxwell stress \citep{Balbus1991}. In addition to the dynamics of angular momentum transport, magnetic fields also drive other high energy processes in astrophysical disks, including the poloidal collimation of jets resulting from fields in the disk. Continued observational discoveries of high energy emission and jets from accretion flows around black hole engines, in particular recent observational evidence of weak, large-scale magnetic fields in disks \citep{akiyama2021}, highlight the fundamental role of magnetic fields in explaining many of the most luminous sources in the universe.
In this letter, we examine whether magnetic fields and the resulting currents from a flow-driven instability {such as MRI} could directly provide sites for fast magnetic reconnection in accretion flows.

Magnetic reconnection, the rearrangement of magnetic field topology, is a ubiquitous process in nature, which energizes magnetically-dominated plasmas such as solar flares, and is being extensively studied through observations \citep{Fox2016}, experiments \citep[][and references therein]{zweibel2009,ji2020major}, as well as for space application \citep{ebrahimi_2020}. Reconnection also plays a critical role in disk corona regions \citep[][and references therein]{Uzdensky_2011} and the creation of flux ropes. 
In the context of MRI turbulence in shearing box geometry, channel flows \citep{balbus_hawley_1992} have been found to be subject to parasitic instabilities such as magnetic tearing-mode reconnection \citep{GoodmanXu1994,Latter_2009}. Reconnection and secondary tearing instabilities have also been investigated using axisymmetic MRI simulations \citep{Dorland} and in the the context of MRI experiments \citep{Liu_goodman_Ji_2006}. However, studies of magnetic reconnection in flow-dominated accreting systems still remain limited. 
The role of turbulent reconnection {in accretion disks has been investigated using shearing box simulations \citep{kadowaki_2018}, and} has also been suggested to be responsible for the radio and gamma-ray emission from accreting black holes \citep{Singh_2015, Ripperda_2020}. 
Joule heating and X-ray luminosity have also been associated with the energy release due to magnetic reconnection in the innermost plunging region of black hole accretion disks \citep{Machida_2003}. 
In protoplanetary disks, the location of current sheets and their distribution may also be important for intermittent heating of these disks \citep{Hubbard_2012}.

\begin{deluxetable*}{cccccccc}[tbh!]

    \tabletypesize{\footnotesize}

    \tablecaption{Initial parameters and calculated values for simulated disks \label{tab:params}}

    \tablehead{\colhead{ID \#} & \colhead{\# of toroidal modes} & \colhead{Aspect Ratio} & \colhead{$\eta$} & \colhead{$R_m$} & \colhead{$S$} & \colhead{$\gamma / \Omega_0$} & \colhead{poloidal resolution} \\ 
    \colhead{} & \colhead{$N$} & \colhead{$H/R$} & \colhead{(\si{\elecd})} & \colhead{$HV_{\phi0}/\eta$} & \colhead{$LV_a/\eta$} & \colhead{} & \colhead{$r \cross z$ polynomial deg.} } 

    \startdata
        \newtag{P1}{sim:P1} & 1 & 0.44 & 5 & \num{2.53E+03} & \num{2.81e2} & 0.259 & $40 \cross 40$ ($80 \cross 80$) deg. 5 \\
        \newtag{P2}{sim:P2} & 1 & 0.44 & 2 & \num{6.32E+03} & \num{6.59e2} & 0.282 & $40 \cross 40$ deg. 5 \\
        \newtag{P3}{sim:P3} & 1 & 0.44 & 1 & \num{1.26E+04} & \num{1.11e3} & 0.291 & $40 \cross 40$ ($80 \cross 60$) deg. 5 \\
        \newtag{P4}{sim:P4} & 1 & 0.44 & 0.5 & \num{2.53E+04} & \num{2.66e3} & 0.301 & $40 \cross 40$ ($80 \cross 80$) deg. 6 \\
        \newtag{P5}{sim:P5} & 1 & 0.44 & 0.2 & \num{6.32E+04} & \num{5.12e3} & 0.307 & $80 \cross 40$ deg. 6 \\
        \newtag{P6}{sim:P6} & 1 & 0.44 & 0.1 & \num{1.26E+05} & \num{1.02e4} & 0.310 & $80 \cross 60$ deg. 6 \\
        \newtag{O1}{sim:O1} & 1 & 2 & 5 & \num{5.06E+03} & \num{2.90e2} & 0.255 & $40 \cross 40$ deg. 5 \\
        \newtag{O2}{sim:O2} & 1 & 2 & 2 & \num{1.26E+04} & \num{7.80e2} & 0.280 & $40 \cross 40$ deg. 5 \\
        \newtag{O3}{sim:O3} & 1 & 2 & 1 & \num{2.53E+04} & \num{1.26e3} & 0.296 & $40 \cross 40$ ($80 \cross 80$)  deg. 5 \\
        \newtag{O4}{sim:O4} & 1 & 2 & 0.5 & \num{5.06E+04} & \num{2.36e3} & 0.297 & $40 \cross 40$ {($80 \cross 80$)} deg. 6 \\
        \newtag{O5}{sim:O5} & 1 & 2 & 0.2 & \num{1.26E+05} & \num{5.37e3} & 0.304 & $80 \cross 80$ {($120 \cross 120$)} deg. 6 \\
        \newtag{O6}{sim:O6} & 1 & 2 & 0.1 & \num{2.53E+05} & \num{1.12e4} & 0.310 & $80 \cross 80$ deg. 6 \\
        \newtag{O7}{sim:O7} & 1 & 2 & 0.05 & \num{5.06E+05} & \num{1.40e4} & 0.318 & $120 \cross 120$ deg. 6 \\
         & 86 (3D) & 2 & 2 & \num{1.01E+04} & \num{6.24e2} & 0.237 & $40 \cross 40$ deg. 4 \\
    \enddata
    
    \tablecomments{Local $S$ is computed shortly after current sheet formation, prior to plasmoid formation. In the 3D case, $S$ was computed at $t=\SI{.32}{ms}$. {In the resolution column, the parentheses indicate the higher resolution used to verify the results.}}

\end{deluxetable*}

In these flow-dominated systems, the question arises whether fluctuations from flow-driven instabilities could directly cause spontaneous plasmoid reconnection.
Magnetic reconnection is characterized by oppositely directed field lines, separated by a current layer, pinched into an X-point {\citep{petschek_1964}} along with a magnetic island, or plasmoid, formation. When the local Lundquist number, defined as $S = LV_a / \eta$, where $L$ is the length of the local current sheet, $V_a=B_0/\sqrt{\mu_0 \rho}$ is the Alfv\'en speed based on the local reconnecting magnetic field strength $B_0$, and $\eta$ is the magnetic diffusivity, becomes sufficiently large, on the order of \num{e4}, thin current sheets are subject to plasmoid instability, or plasmoid reconnection, in which they break up into multiple plasmoids and X-points {\citep{Biskmap_1986,tajima_shibata_1997, Loureiro2007}}.

In this letter we examine the underlying physics of accretion flows and whether there are reconnection sites generated by the primary MRI that could trigger {fast} plasmoid reconnection. We employ a global resistive MHD disk model, a hollow cylinder threaded by a weak vertical magnetic field.
In particular, we focus on the formation and evolution of current sheets under MRI, and their involvement in reconnection events, for low and high Lundquist numbers and in two and three dimensions. 
To identify the reconnection sites and X-points in 2D and 3D, we utilize Poincar\'e plots in the poloidal plane (the intersection of field lines with the $rz$ plane as the field is followed around azimuthally), as well as in the toroidal $xy$ planes.
Through this field line tracing, we find that 2D plasmoids (and 3D magnetic islands) are formed both through driven reconnection (i.e. via  merging current sheets from MRI turbulent motions), as well as through spontaneous plasmoid reconnection due to current sheet instabilities driven by MRI.

\section{Simulation Setup}

To simulate an accretion disk subject to MRI, we use the NIMROD (Non-Ideal Magnetohydrodybamics with Rotation, an Open Discussion project) code \citep{SOVINEC2004355}, which solves the three-dimensional nonlinear extended MHD equations {in toroidal axisymmetric, cylindrical, or Cartesian slab geometry using a pseudo-spectral method}. For our MRI simulations presented in this paper, we employ an MHD model with explicit resistivity and with a similar simulation setup as described in \citet{Ebrahimi_2011_hall}. 
\deleted{Refer to section 2 of that paper for a more detailed description of NIMROD and the MHD model.} 

\begin{figure*}[bt!]
    \centering
    \gridline{\fig{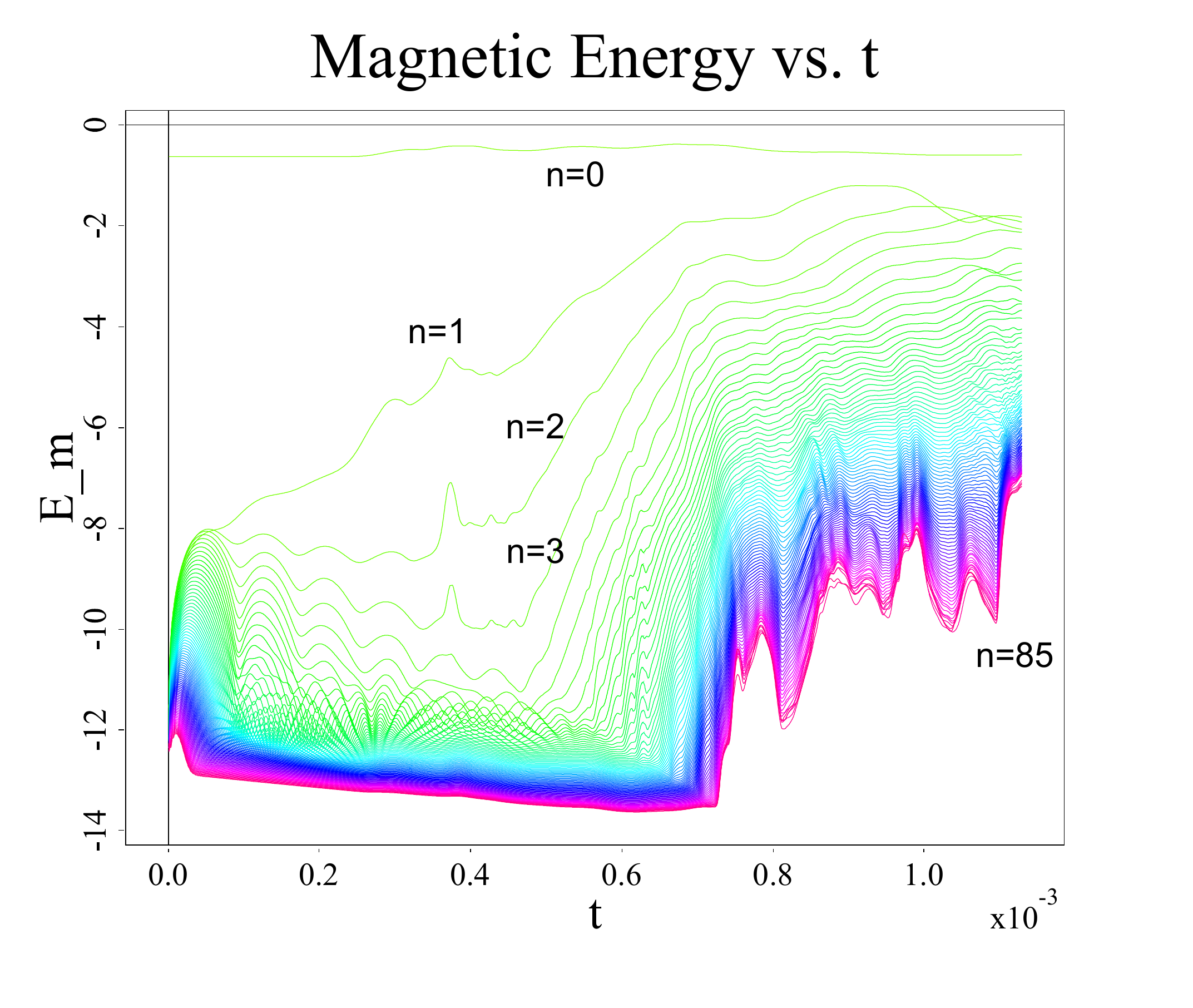}{.45\textwidth}{(a)}
    \fig{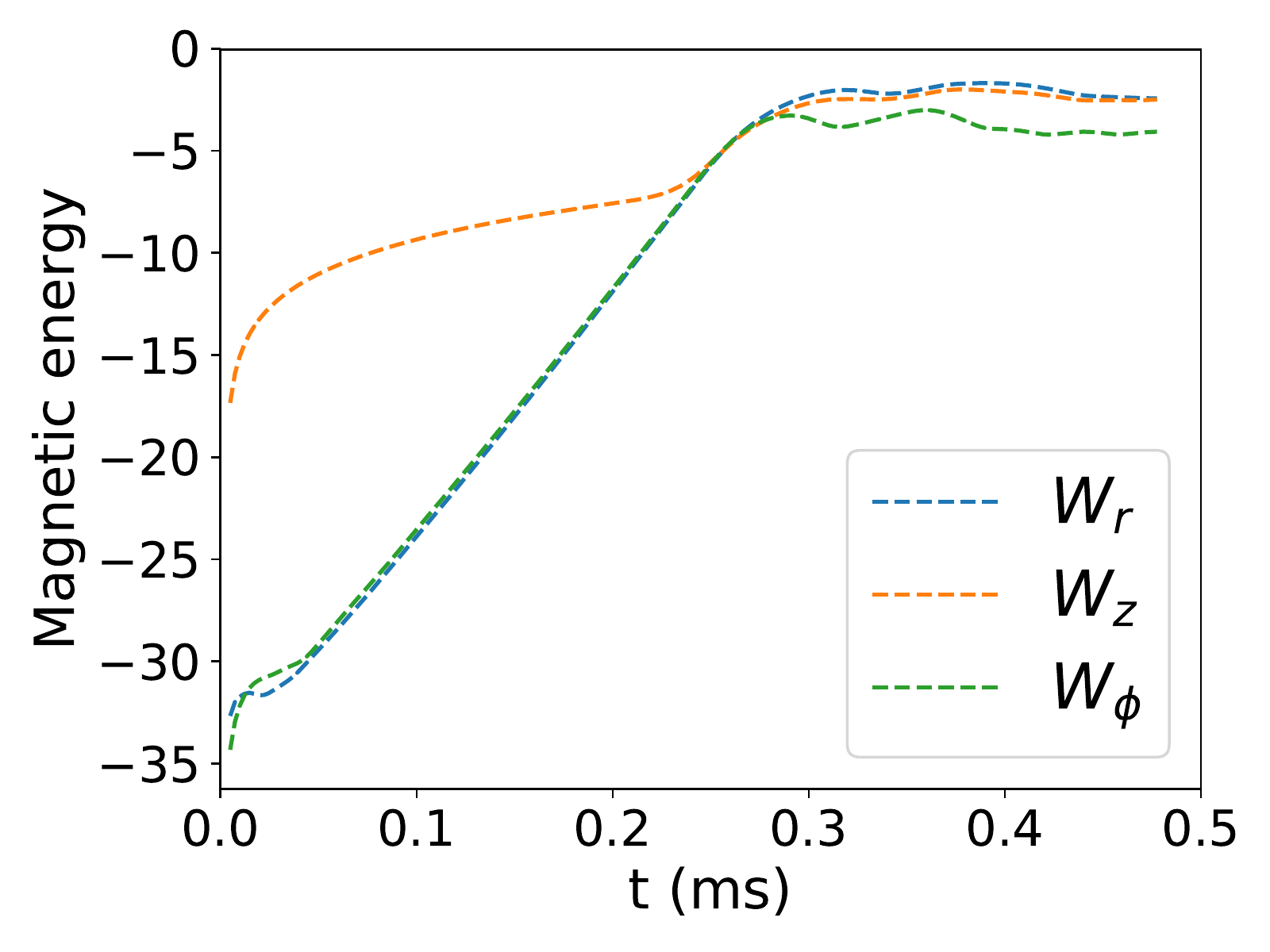}{.45\textwidth}{(b)}}
    
    \gridline{\fig{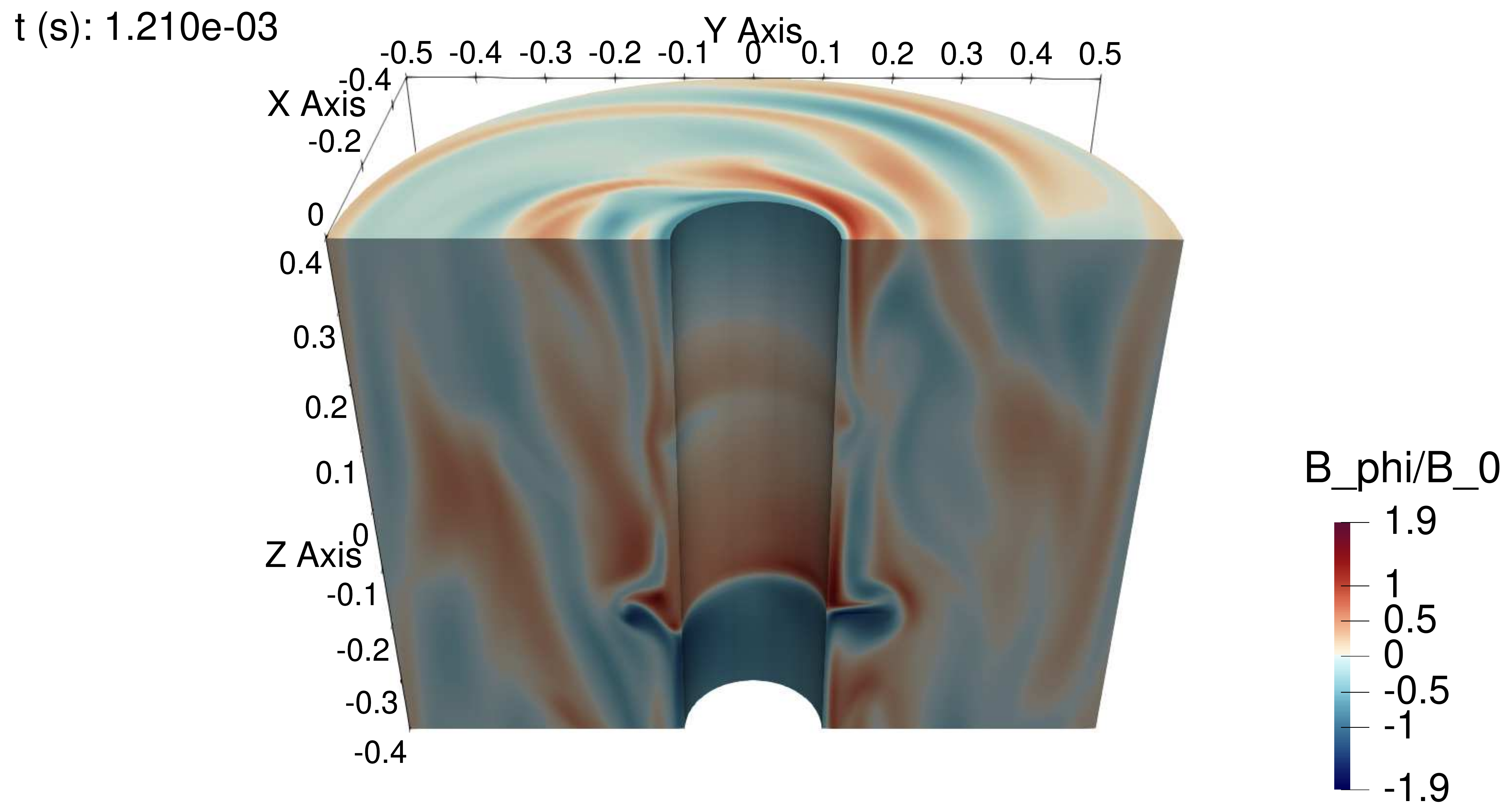}{.5\textwidth}{(c)}
    \fig{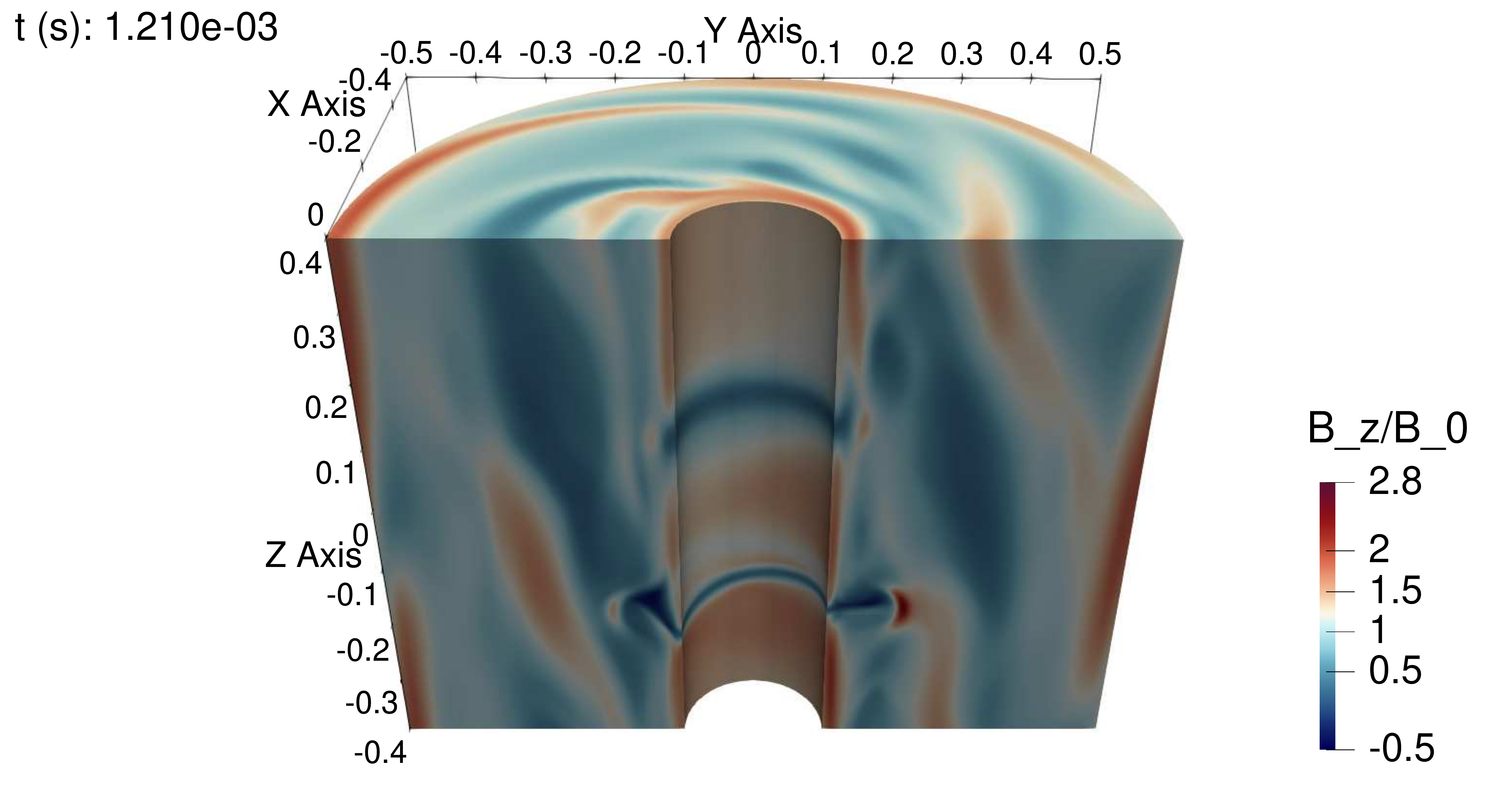}{.5\textwidth}{(d)}} 
    
    \caption{The log of the total magnetic energy with respect to time for each toroidal Fourier mode is given in (a) for the 3D case. The top curve is the $n=0$ mode, the next one down is the $n=1$, and so on. The {log of} volume averaged magnetic energy $W_i = 1/2\mu_0 V \int (B_i-B_{i0})^2 \dd{V}$, where $B_{i0}$ is the value of the field component at $t=0$, is plotted in (b) early in time for the $n=0$ mode, broken down by component. A poloidal cut of the cylinder late in time, colored by $B_\phi$, is displayed in (c). A poloidal cut of the cylinder late in time, colored by $B_z$, is similarly displayed in (d).}
    \label{fig:3D_E}
\end{figure*}

We model the accretion disk as an unstratified Keplarian cylinder with an impenetrable hollow center where the supermassive object would normally sit (see figure \ref{fig:3D_E}c and \ref{fig:3D_E}d for reference on the general geometry of our disk). We do not include gravity in our model, as we are only interested in the MRI and its associated plasma physics, and instead balance the momentum equation with a pressure gradient $\grad p = \rho V_\phi^2 / r$. We simulate a deuterium plasma with a number density of \SI{e19}{m^{-3}} (all the quantities in NIMROD are in physical SI units). The relevant magnetic diffusivities $\eta$ used in our simulations are given in table \ref{tab:params}. Across our simulations we cover a range of magnetic Reynolds numbers, from \num{e3}-\num{e5} (see table \ref{tab:params}), defined as $R_m = HV_{\phi0}/\eta$, where $H = z_2 - z_1$ is the cylinder height and $V_{\phi 0}$ is the initial toroidal flow at the inner boundary. The local Lundquist numbers $S$ have also been computed shortly after current sheet formation, using $B_r$ {(radial magnetic field perturbation fully generated by MRI)} near the sheets for the Alfv\'en speed, and included in the table. The magnetic Prandtl number, defined as $P_m = \nu / \eta$ where $\nu$ is the viscosity, was kept at unity across all simulations.

We use a finite element discretization in the poloidal plane ($rz$). We give the numerical resolution of our simulations in table \ref{tab:params}, which we define as the number of elements in $r$ by $z$, along with the polynomial degree.
We use a spectral, discrete Fourier transform in the toroidal ($\phi$) direction. Variables, including the physical fields, flow velocity and magnetic field, thus are represented in standard cylindrical coordinates ($r$, $\phi$, $z$) and have the following form:
\[ f(r,\phi, z, t) = f_{0}(r, z, t) + \sum_{n=1}^{N-1} \left[ f_n e^{in\phi} + f_n^* e^{-in\phi} \right] \]
where $N$ is the total number of toroidal modes included, and $f_n$ are complex functions of $r$, $z$, and $t$. To study reconnection physics, we performed several 2D, axisymmetric simulations with varying explicit resitivities\footnote{MRI has been studied under varying $\eta$ before \citep[see][]{Fromang_2007, Fromang_Stone_2009}, though not in the context of reconnection}, achieved by only evolving the $n=0$ mode ($N=1$). To study the impact of 3D, non-axisymmetric effects, we carried out one simulation with $N=86$ toroidal modes.
The toroidal plasma flow at the boundaries are specified to drive flow, and a no-slip condition is used for the poloidal flow. The cylinder walls are perfectly conducting. The toroidal and vertical ($z$) directions are periodic.

{We should note that, in order to verify our results, we took a subset of our 2D simulations with varying resistivities and doubled the poloidal resolution (see table \ref{tab:params}), and we found no significant changes in our results. Due to time and resource constraints, we did not check our 3D simulation in the same fashion for the poloidal resolution. Thus, while the 3D results should be reliable early in time in the 2D regime, the number of reconnection sites might require further verification later in time at higher resolution. As we are interested in the global nature of reconnection sites and for the sake of completeness, we will show some of the interesting results from the 3D simulation at the end of this paper. A more thorough study of reconnection in the turbulent regime with higher resolution will be investigated in a future work.}

To trigger MRI in our simulations, we seed the disk with a uniform magnetic field aligned in the $z$ direction, and impose an initial Keplarian toroidal flow of the form $V_\phi \propto r^{-1/2}$. The initial profiles, including the magnitude of the magnetic field, in the simulations are chosen to be unstable to MRI in the MHD regime {(with $B_0 = \num{0.001}{T}$)}, and have been benchmarked with the local WKB dispersion relation to produce a sufficiently large growth rate \citep[see][fig. 6]{Ebrahimi_2011_hall}. The MRI linear growth rate $\gamma$, normalized to the initial inner angular velocity $\Omega_0$, has also been computed for each simulation and included in table \ref{tab:params}.

For our 2D study, we examined two types of cylinders: one with dimensions $\SI{0.1}{m} \leq r \leq \SI{0.5}{m}$ and $-\SI{0.4}{m} \leq z \leq \SI{0.4}{m}$, and the other with $\SI{0.1}{m} \leq r \leq \SI{1}{m}$ and $-\SI{0.2}{m} \leq z \leq \SI{0.2}{m}$. Defining the aspect ratio to be $H/(r_2-r_1)$, the former has a ratio of \num{2.0} and the latter a ratio of \num{0.44}. We shall refer to these two cylinders by their aspect ratio going forward. For our 3D disk, we use aspect ratio 2.0.

\section{Results and Analysis}

\begin{figure}[hbt!]
    \centering
    \gridline{\fig{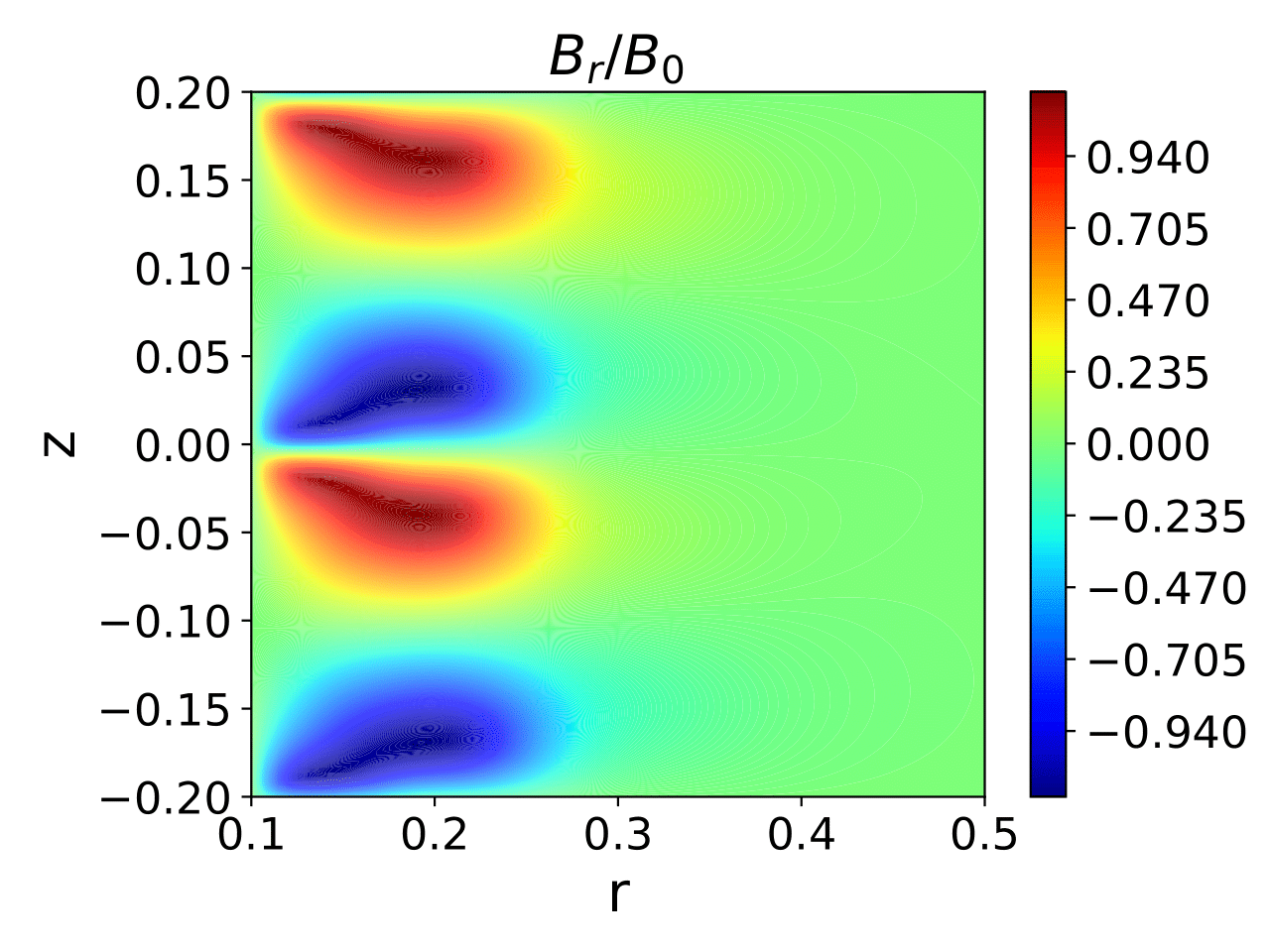}{\linewidth}{(a)}}
    \gridline{\fig{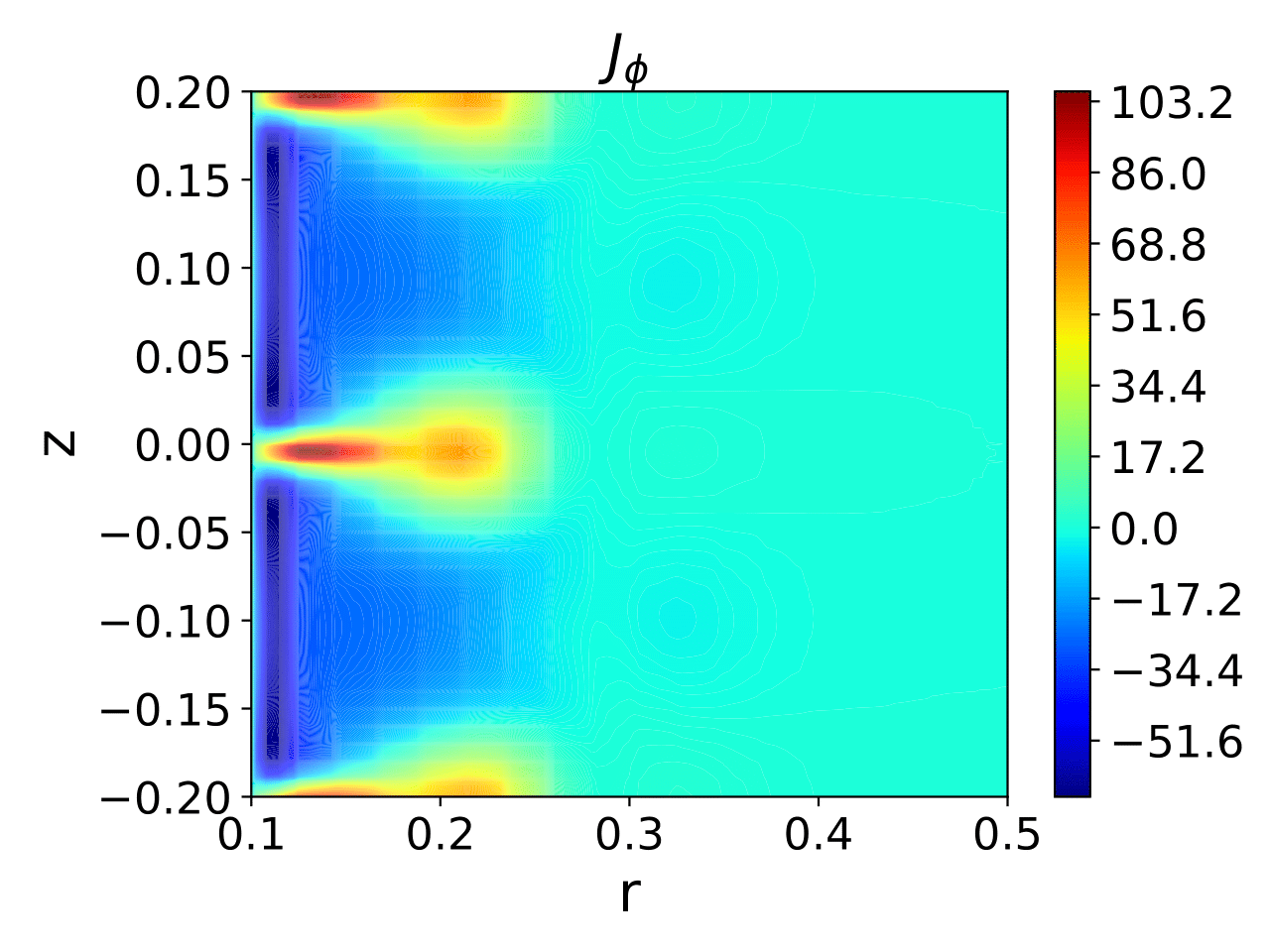}{\linewidth}{(b)}}

    \caption{The radial component of the magnetic field, $B_r$, is shown in (a), {normalized by $B_0$} early in time for {disk \ref*{sim:P1}}. The corresponding toroidal current density $J_\phi$ is shown in (b), in units of kilo-amps per meter squared (we will continue to use these units for current density in future plots unless otherwise stated).} 
    
    \label{fig:elecd5_pan}
\end{figure}

We began our study by performing the full 3D simulation of standard MRI in global cylindrical geometry as described above. 
Figure \ref{fig:3D_E}a shows the evolution of total magnetic energy in the 3D simulation, separated by mode, with respect to time. Figure \ref{fig:3D_E}b is the energy of each component of the axisymmetric $n=0$ mode early in time, where the mean initial $B_{z0}$ has been subtracted from $B_z(t)$. 
The axisymmetric MRI $n=0$ mode grows exponentially (with $\gamma/\Omega_0 = \num{.237}$) and saturates relatively quickly, at time $t=\SI{.3}{ms}$. Many of the higher, non-axisymmetric modes are nonlinearly triggered and saturate around $t=\SI{.7}{ms}$. As is seen from the modal energy and spectrum in figure \ref{fig:3D_E}a, this 3D simulation is well resolved in the toroidal direction. 

\begin{figure*}[bt!]
    \centering
    \gridline{\fig{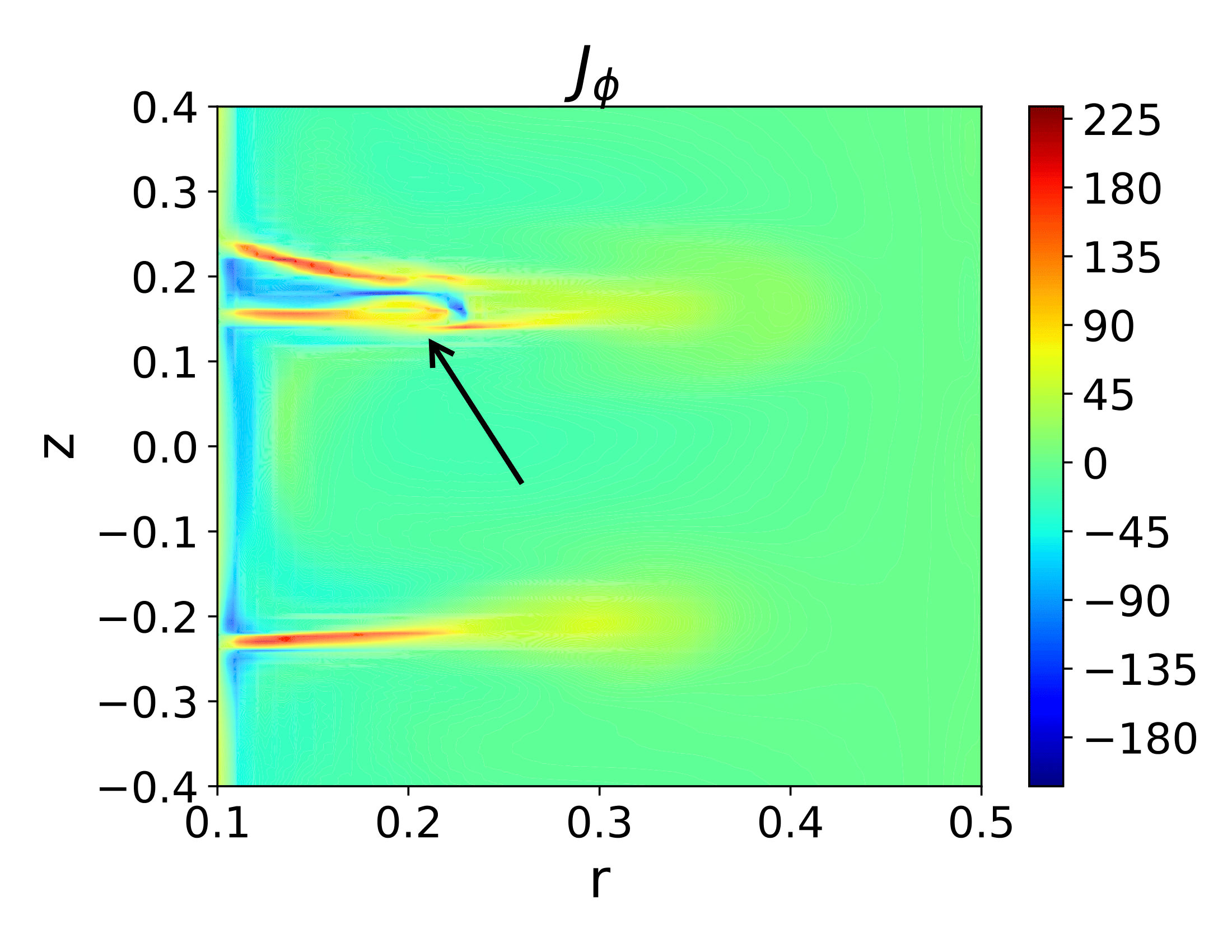}{.45\textwidth}{(a)}
    \fig{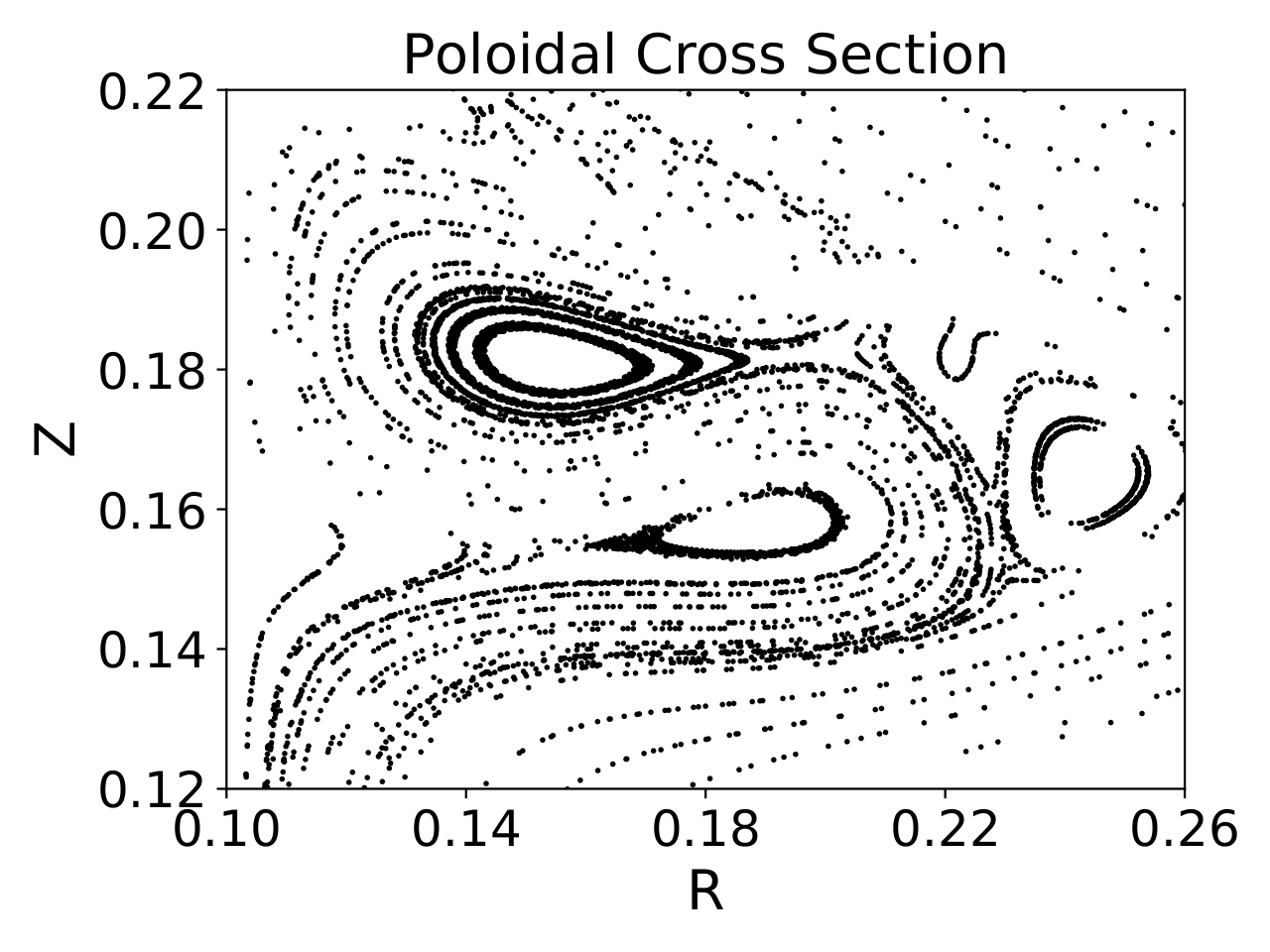}{.45\textwidth}{(b)}}
    
    \gridline{\fig{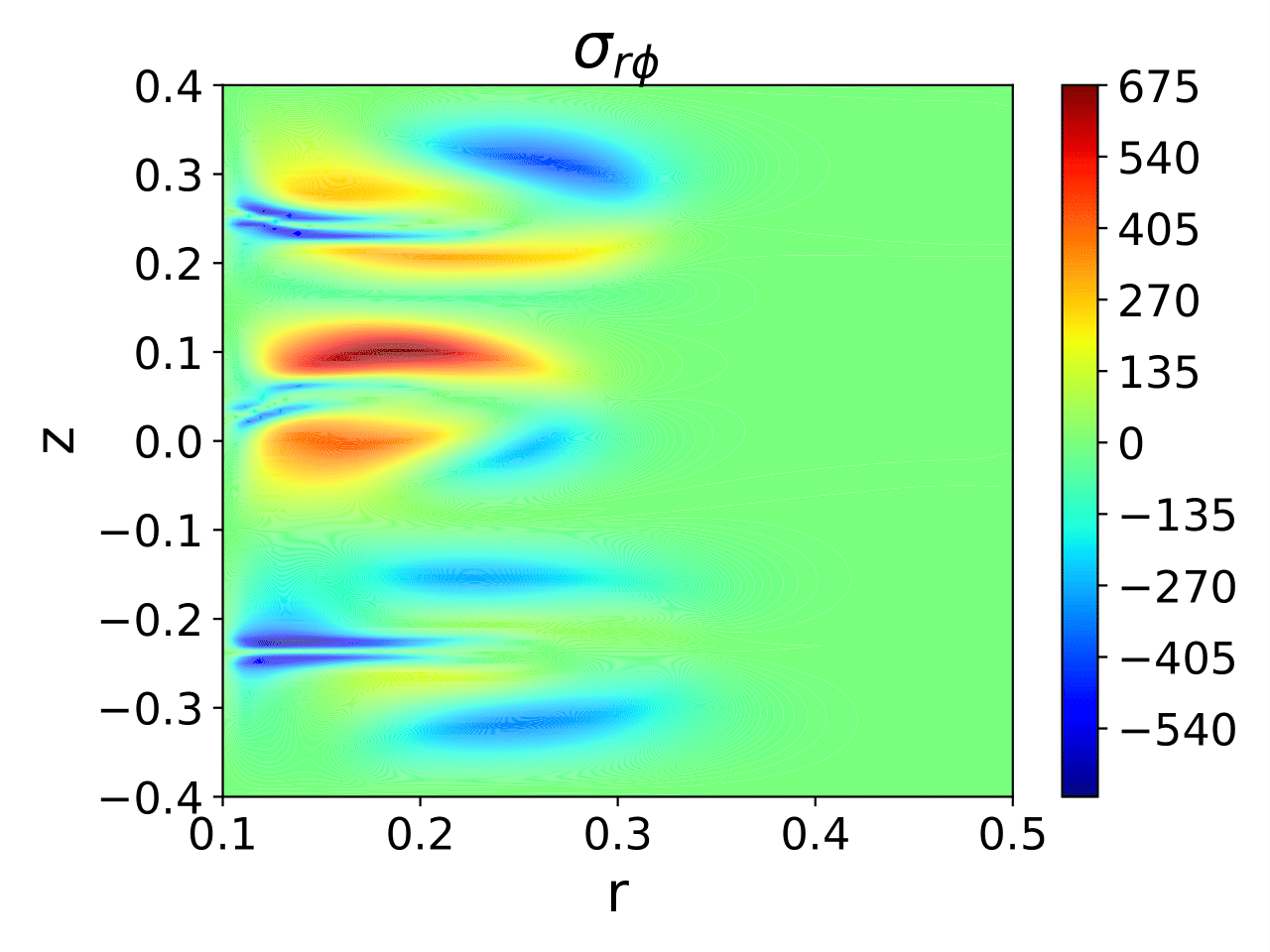}{.45\textwidth}{(c)}
    \fig{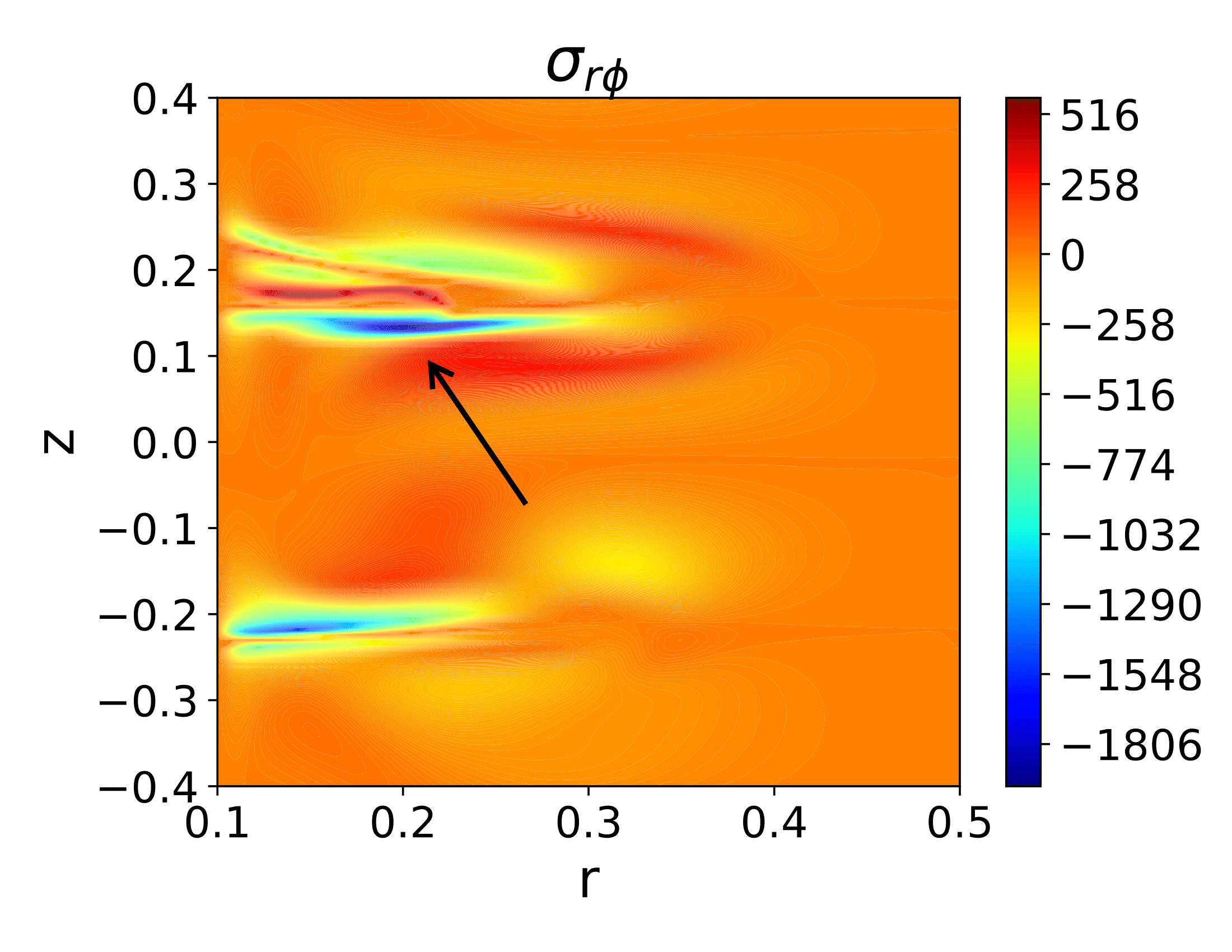}{.45\textwidth}{(d)}}
    
    \caption{The current density $J_\phi$ during a merging event for the 3D case at time $t=\SI{0.385}{ms}$ is shown in (a). The black arrow points to the merging region. The corresponding Poincar\'e plot of the magnetic field is given in (b), zoomed around the region with the two upper current sheets shown by arrow. The Maxwell stress $\sigma_{r\phi}$ is show in (c) prior to merging, at time $t=\SI{.320}{ms}$, while (d) is the Maxwell stress at the time shown in (a) and (b), where the black arrow points to the region of merging.}
    \label{fig:mergers}
\end{figure*}

For reference, an image of the top and inner plane of the cylinder is shown in figures \ref{fig:3D_E}c and \ref{fig:3D_E}d, late in time, at which point the system has become turbulent from higher modes saturating. Note that the magnitude of $B_\phi$ {has become twice as strong as the starting vertical field in some spots, and the magnitude of $B_z$ has increased to almost triple it's starting value}. This is a clear indication of dynamo action as a result of MRI induced turbulence \citep{brandenburg_1995, Rincon_2007, Lesur_2008,Ebrahimi_2009, ebrahimi_blackman_2016, bhat_ebrahimi_blackman_2016}. In particular, the radially alternating large-scale magnetic fields (as seen in figures \ref{fig:3D_E}c and \ref{fig:3D_E}d) were also shown through quasilinear theoretical and numerical analysis \citep{ebrahimi_blackman_2016}.

Before the 3D non-axisymmetric modes begin to saturate at around $t=\SI{0.7}{ms}$, we observed some interesting 2D axisymmetric dynamics {(see figure \ref{fig:3D_E}a)}. Current sheets start forming near the end of the growth phase of the $n=0$ mode, around $t=\SI{0.25}{ms}$. Interestingly, at around $t = \SI{0.35}{ms}$, there is a slight burst, followed by a quick drop, in the non-axisymmetric magnetic energy, which resembles a magnetic reconnection event. In order to understand these dynamics, we performed a systematic 2D numerical study, as described in the previous section. The evolution of magnetic energy in the 2D axisymmetric simulations look similar to that of figure \ref{fig:3D_E}b, as one might expect.

Below, we first discuss the process of current sheet formation during our axisymmetric simulations. 
The dynamics and the nonlinear evolution of current sheets are then discussed for simulations at low $S$ when current sheet merge and at high $S$ when current sheets break due to plasmoid instability.
Finally, we show that 3D turbulent effects vertically stretch and dissipate formed current sheets, and produce large magnetic islands in the outer regions of the disk and small islands in the inner region of the toroidal plane.

\subsection{Current sheet formation due to axisymmetric MRI}

In both our 2D simulations and 3D simulation when the $n=0$ axisymmetric mode is dominant, we observe the formation of toroidal current sheets. As MRI sets in, a radial perturbation develops in the otherwise vertical magnetic field, as seen in figure \ref{fig:elecd5_pan}a. According to Amp\'ere's Law, toroidal current density is given as
$ \mu_0 J_\phi = \pdv*{B_r}{z} - \pdv*{B_z}{r} $
In the thin space between the positive and negative regions of $B_r$ (radial magnetic field perturbation fully generated by MRI), the first term on the right grows very large, causing a relatively strong toroidal current $J_\phi$ to form, as shown in figure \ref{fig:elecd5_pan}b. These Harris type current sheets localized vertically as the result of large gradient in $B_r$ appear near the end of the growth phase for the $n=0$ mode. We again emphasize that these current sheets are axisymmetric and form in all our 2D and 3D cases (when $n=0$ is dominant), however their dynamics vary based on the Lundquist number and the number of significant toroidal modes, as described below. 
{We should also note that in 2D, current sheets are formed as the result of a steep vertical gradient in large-scale axisymmetric MRI perturbation, and the number of current sheets formed here depends on the vertical wave number of the primary MRI mode (and thus the aspect ratio of the system).}

\subsection{Current Sheet Mergers at low S}

As we investigated the dynamics of current sheets formed around the time of the burst in figure \ref{fig:3D_E}a, we noted that one of the current sheets moves towards another, colliding with it and combining into one current sheet. See figure \ref{fig:mergers}a for a snapshot of this merging in progress. This is, in part, a product of the attractive Lorentz force felt by two parallel sheets of current moving in the same direction. We observed merging in all of our simulated disks. For the 3D case and 2D cases with $S > \num{3e2}$ ($\eta < \SI{5}{\elecd}$), this merging causes plasmoid formation (closed magnetic loops) due to pinching and compression of magnetic field, as seen in figure \ref{fig:mergers}b (the spot where the magnetic fields are pinched together is often referred to as an X-point). Thus, the merging of current sheets acts as a catalyst for magnetic reconnection.

We can determine what effect the mergers have on angular momentum transport by calculating the $r\phi$ component of the Maxwell stress tensor: $\sigma_{r\phi} = B_r B_\phi / \mu_0$. Negative stress corresponds to outward momentum transport \citep[see][]{Balbus1998}. During the merging process, the magnitude of the Maxwell stress increases significantly, by roughly three times it's starting magnitude, in the merging region (see figures \ref{fig:mergers}c and \ref{fig:mergers}d). This local burst in momentum transport is consistent with the findings of \citet{Ebrahimi_2011_mom_tran}. There, the radially localized current sheets caused outward momentum transport from the tearing modes in a disk.

\subsection{Onset of Plasmoid Instability at high S}

\begin{figure}[h!]
    \centering
    \gridline{\fig{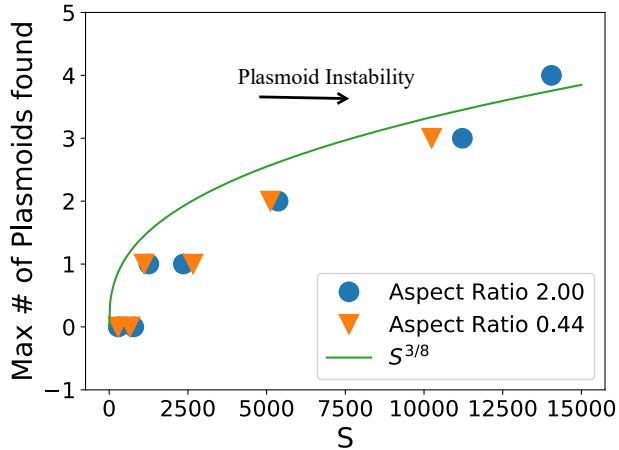}{\linewidth}{(a)}}
    \gridline{\fig{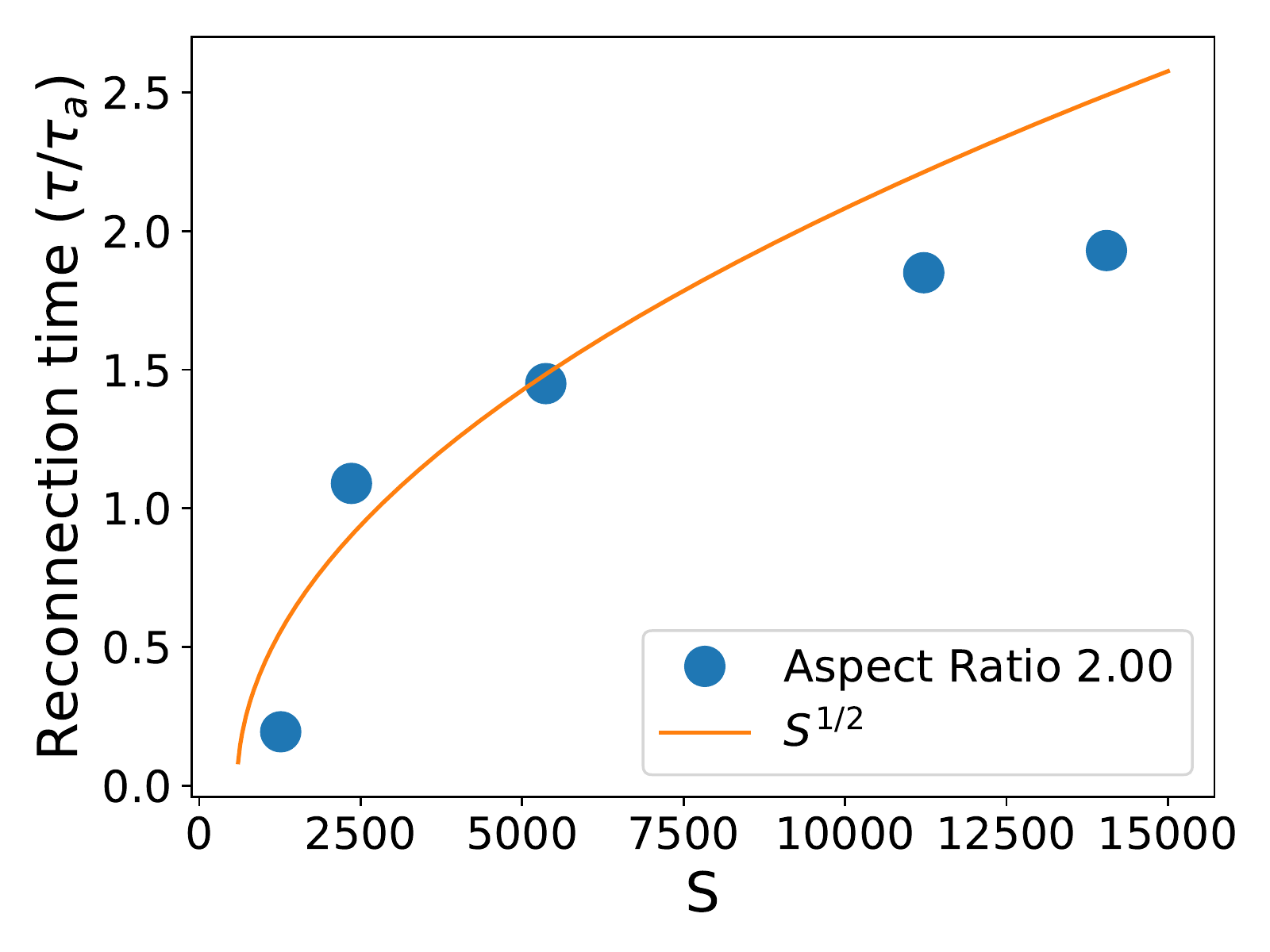}{\linewidth}{(b)}}
    
    \caption{The maximum number of plasmoids in a current sheet from our 2D simulations for a given Lundquist number $S$ with both aspect ratios are shown in (a). {The origin of the black arrow denotes the start of plasmoid instability and points in the direction of increasing $S$.} The green curve is the theoretical scaling $S^{3/8}$, included for comparison {to show  the trend of increasing number of plasmoids with local S}. Note that it is not a rigorous fit, more data is required. {The estimated MRI-driven reconnection times for the taller disk, normalized by the Alfv\'en time ({computed locally as $\tau_A = L/V_A$}) are shown in (b). The orange line is a fit of the S-P theoretical scaling $S^{1/2}$, included for comparison.}}
    \label{fig:plas_vs_S}
\end{figure}

\begin{figure*}[h!]
    \centering
    
    \gridline{\fig{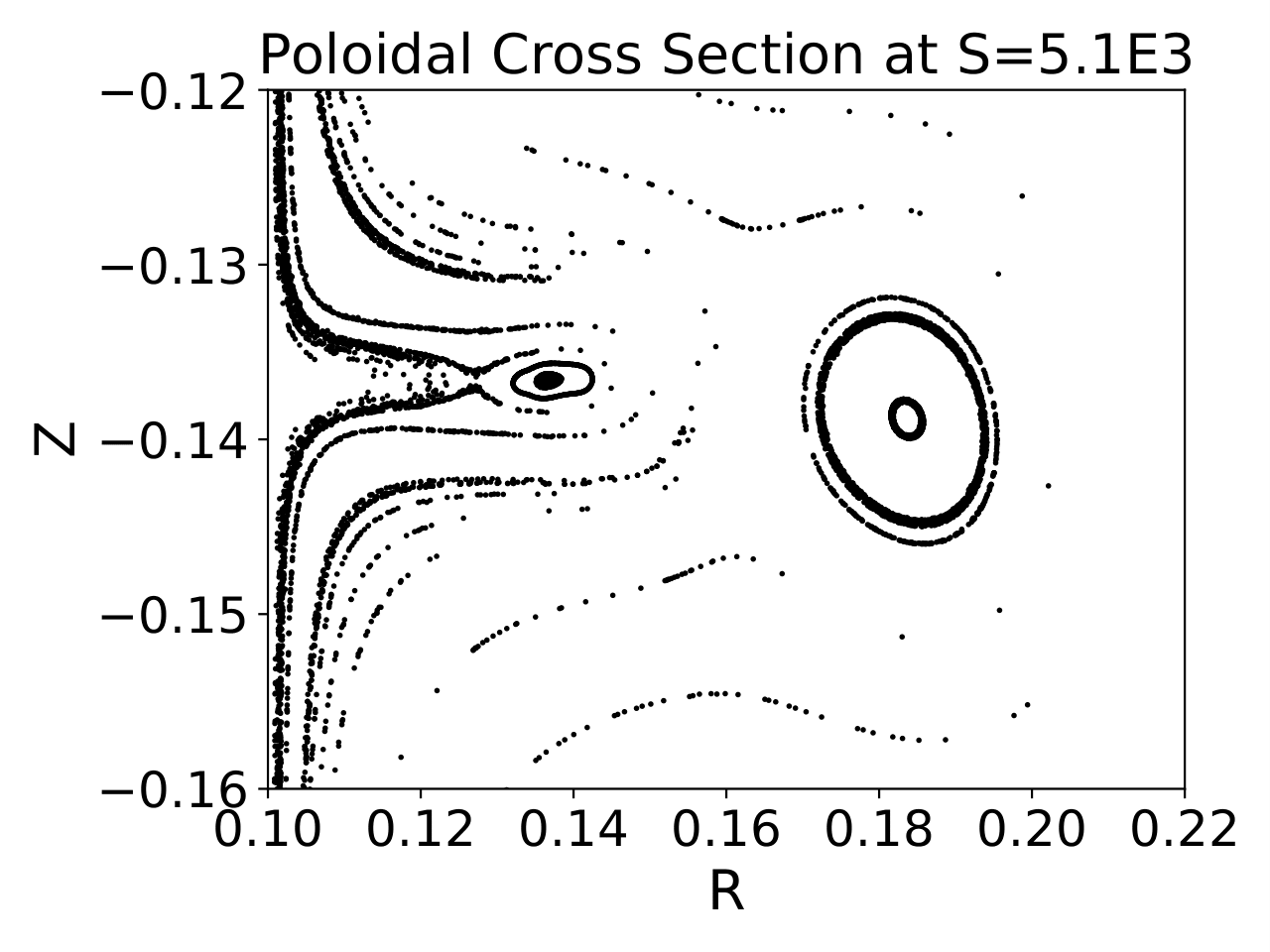}{.45\textwidth}{(a)}
    \fig{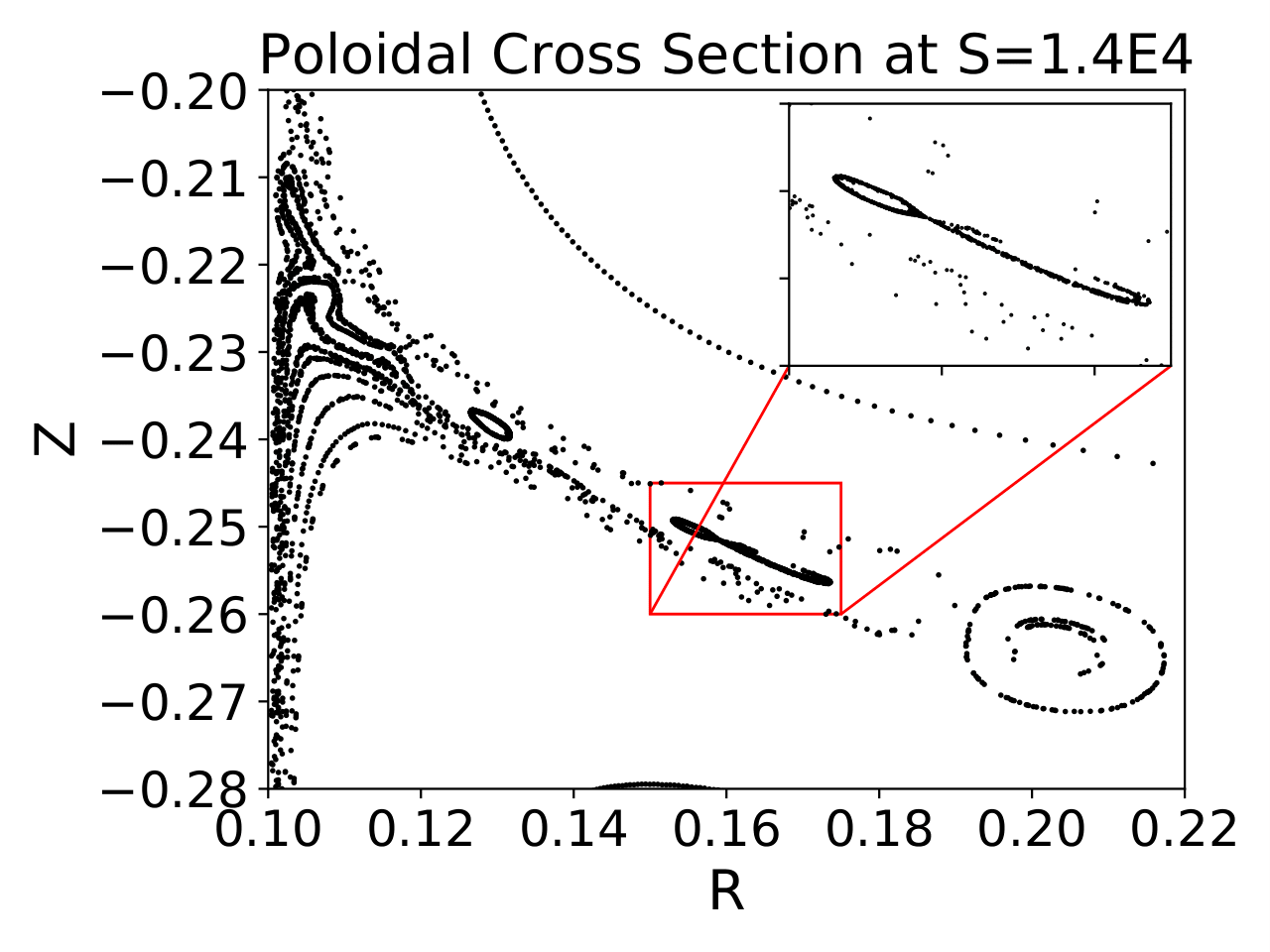}{.45\textwidth}{(b)}}
    
    \gridline{\fig{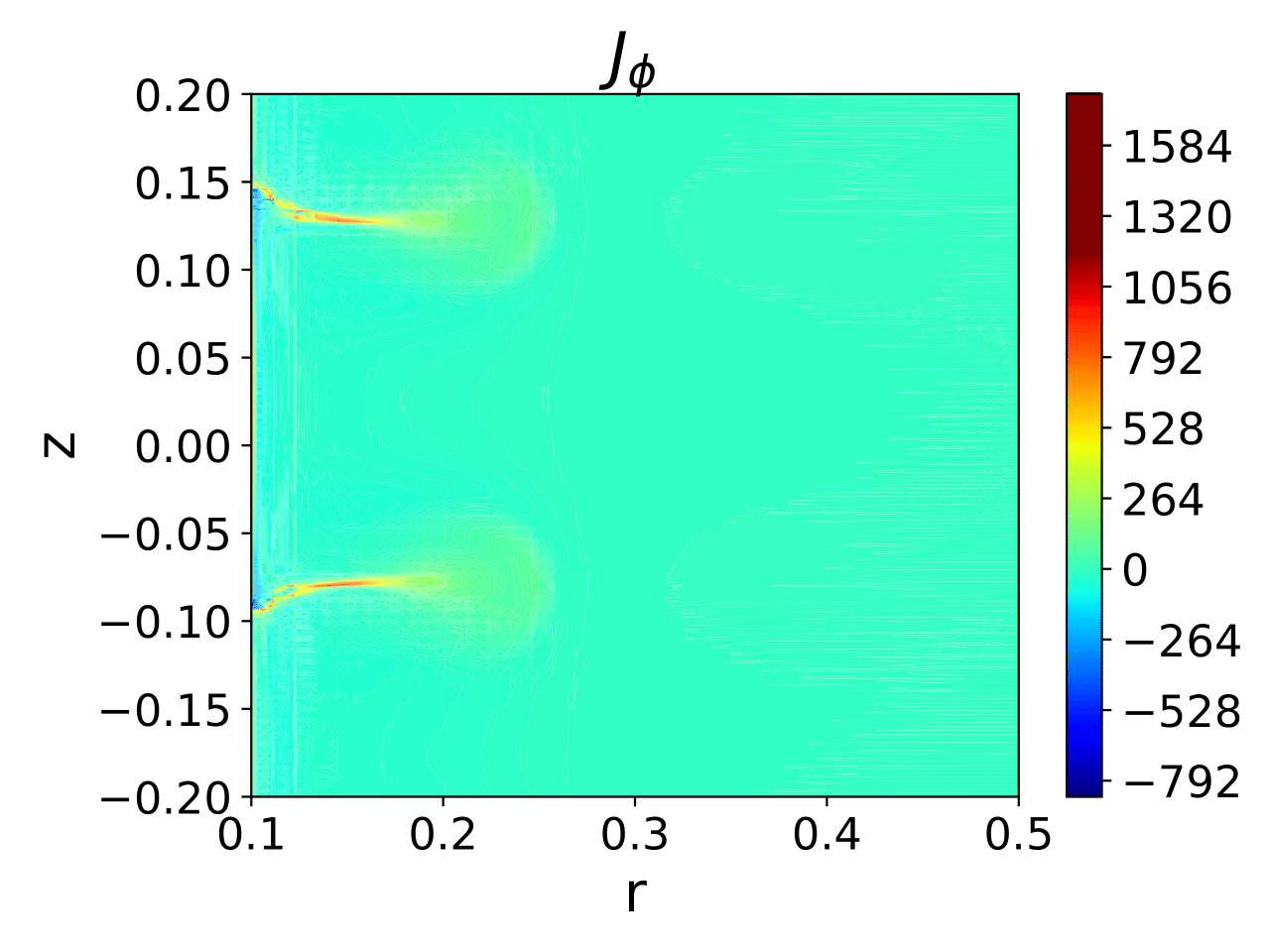}{.45\textwidth}{(c)}
    \fig{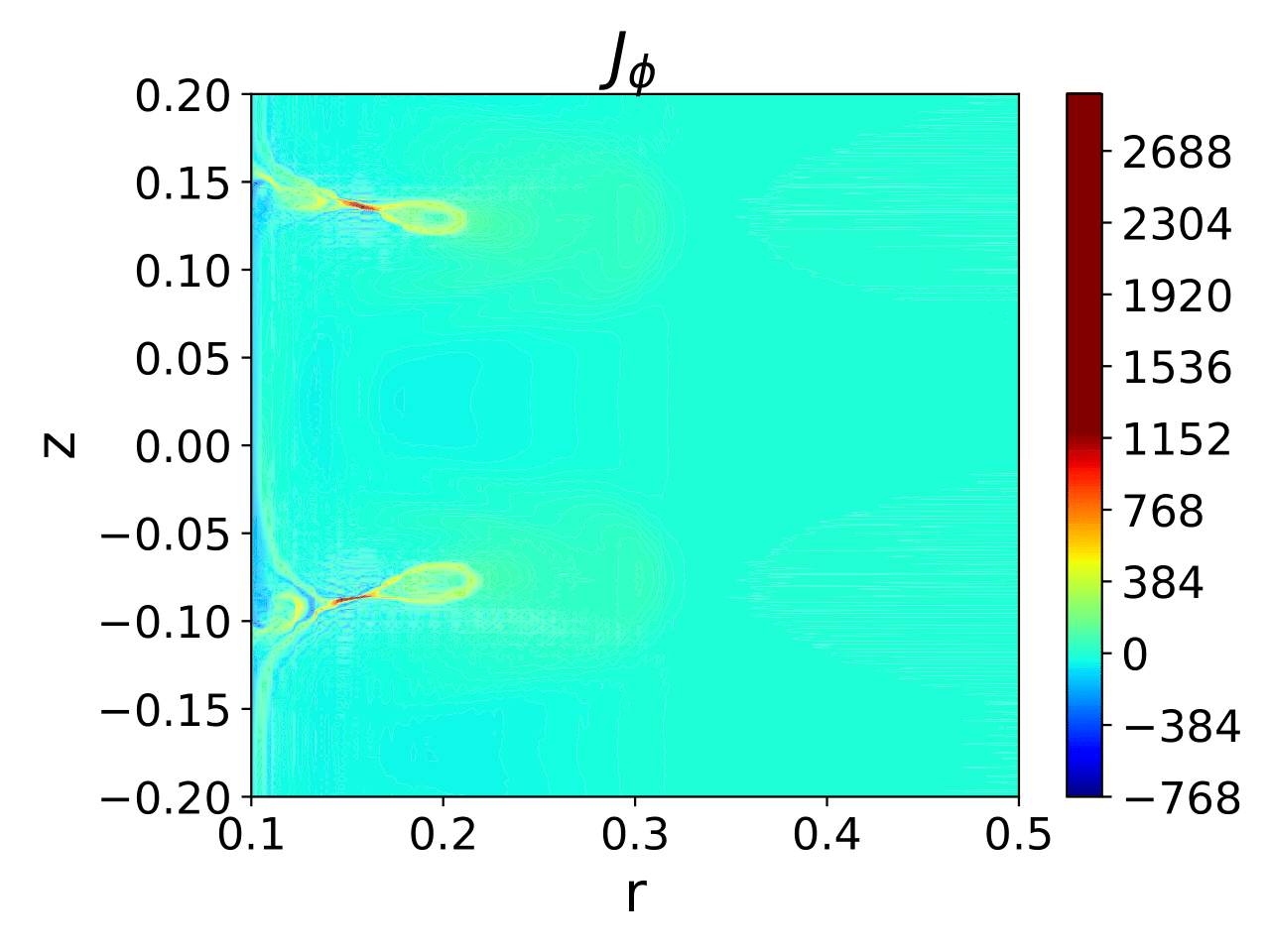}{.45\textwidth}{(d)}}
    
    \gridline{\fig{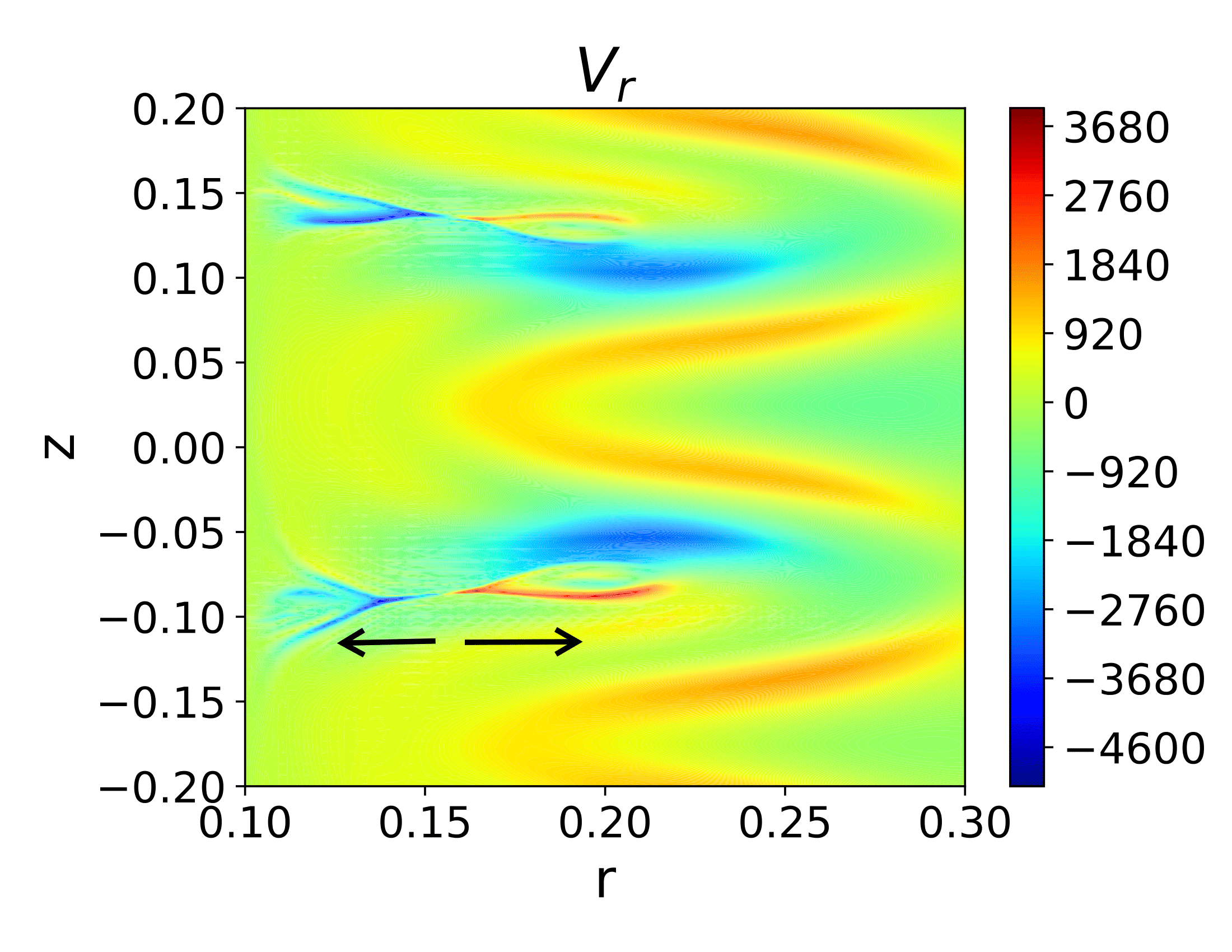}{.45\textwidth}{(e)}
    \fig{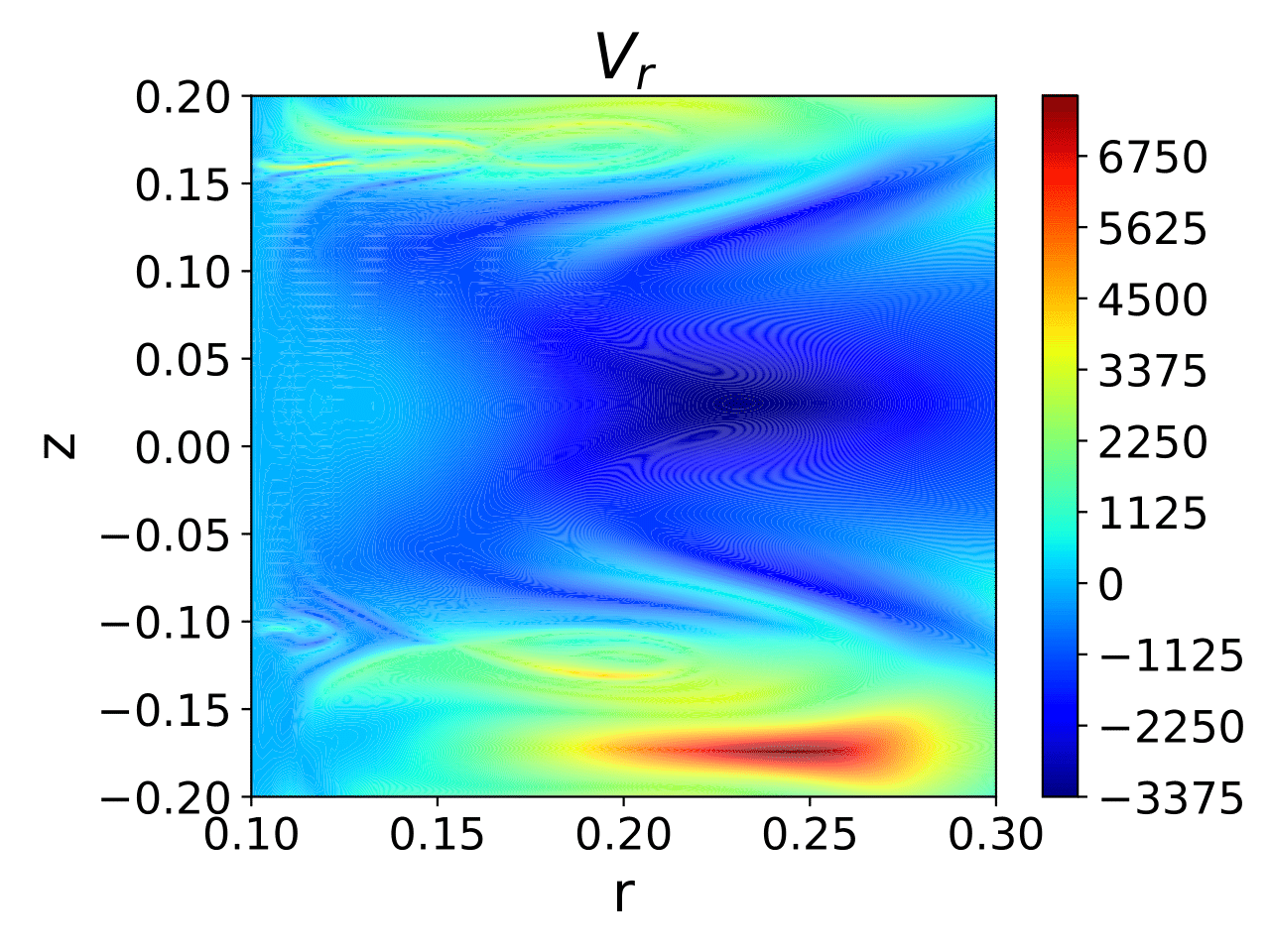}{.45\textwidth}{(f)}}
    
    \caption{(a): Poincar\'e plot of the magnetic field around a current sheet containing two plasmoids {in disk \ref*{sim:P5}}. (b): Magnetic field around a current sheet containing {four} plasmoids {in disk \ref*{sim:O7}}. (c): Toroidal current density just after the current sheets have formed for {disk \ref*{sim:P6}}. (d): Toroidal current density {in disk \ref*{sim:P6} just after plasmoids form}. (e): The radial plasma flow $V_r$ at the time shown in (d). The arrows denote the direction of flow around the X-point in the lower current sheet. (f): The radial plasma flow a few timesteops after the plasmoids appeared.}
    \label{fig:cur_mag_pan}
\end{figure*}

In addition to merging of current sheets and the formation of closed flux regions (driven plasmoid formation) as shown in figure \ref{fig:mergers}b, we also observe additional dynamics at high Lundquist number in the 2D regime.
Spontaneous plasmoids appear in the current sheets. At Lundquist numbers of \num{e3} and up, at least one plasmoid can appear in a current sheet. At \num{5e3} and up, two plasmoids can appear. At {\num{1.4e4}}, we have observed up to {four} plasmoids form in a sheet. See figure \ref{fig:plas_vs_S}a for a plot of the maximum number of plasmoids we observed to form in a current sheet for each 2D simulation. As expected from the theoretical scaling, the number of plasmoids increases with $S$. Note that the scaling of $S^{3/8}$ is only shown for comparison \citep{tajima_shibata_1997}. Figures \ref{fig:cur_mag_pan}a and \ref{fig:cur_mag_pan}b are Poincar\'e plots of the magnetic field when there are two and {four} plasmoids present respectively. Notice that in each case there are multiple X-points, where the magnetic field is getting pinched.
{There is one plasmoid in particular that is larger than the rest, and this is often referred to as the primary, or "monster", plasmoid} {\citep{uzdensky2010fast}}.
Multiple plasmoids with multiple X-points is a sign that the current sheets are undergoing plasmoid instability {\citep{Biskmap_1986, tajima_shibata_1997, shibata_tanuma_2001, Loureiro2007, Bhatteracharjee2009, Cassak2009_SP, Samtaney2009_chains, Huang_Bhattacharjee_2010, Ji2011_phase, Loureiro_2012, Murphy2013_asym, Yu_2014, ebrahimi_2015, Ni_2015, Tenerani_2015, Comisso2016_visco}}.

{We have also estimated the MRI-driven reconnection times for a few of the simulations and included them in figure \ref{fig:plas_vs_S}b. These rates were computed by finding the difference in time between the formation of the X-point and the appearance of the primary plasmoid. As expected, the reconnection time is roughly proportionate to Sweet-Parker \citep{Parker_1957, Sweet_1958} scaling of $S^{1/2}$ before the onset of plasmoid instability.
We should stress that these times are only a rough estimate, however they do show that as $S$ becomes larger the reconnection time starts to become weakly dependent on $S$ \citep{Daughton_2009, Bhatteracharjee2009,Huang_Bhattacharjee_2010, ebrahimi_2015}. As our simulations are global, to further obtain data at higher values of local $S$ shown in figure \ref{fig:plas_vs_S}, much higher poloidal resolution around the 2D MRI-driven current sheets is required.}

{In addition to increased number of plasmoids and faster reconnection rates,} the current sheets themselves become thinner and more unstable at high Lundquist number. Comparing figure \ref{fig:elecd5_pan}b to figure \ref{fig:cur_mag_pan}c, it is clear that the formed current sheets are significantly thinner in the latter plot, at higher Lundquist number. Not long after the current sheets have formed they break into loops as the plasmoids form, as shown in figure \ref{fig:cur_mag_pan}b. Thinner, Sweet-Parker (S-P) type sheets {\citep{Parker_1957, Sweet_1958}} are observed between plasmoids, as the primary current sheet breaks. Breaking current sheets is another sign of plasmoid instability.

Finally, another signature of spontaneous reconnection is outflows at the X-points, and we identify this along the primary current sheet. When magnetic reconnection occurs, magnetic energy is converted into kinetic energy in the plasma, inducing a flow parallel to the plane of the current sheet. As can be seen in figure \ref{fig:cur_mag_pan}e, this does in fact occur, with the plasma flowing radially inward left of the primary X-point and outward right of the X-point. Interestingly, a short while after the appearance of the plasmoids, as seen in figure \ref{fig:cur_mag_pan}f, the flow changes to be fully outward in the current region.
{Further detailed analysis of momentum transport and the reconnection rates at yet higher $R_m$ and $S$ remain for a future work.}

\subsection{Non-axisymmetric effects}

\begin{figure*}[h!]
    \centering
    \gridline{\fig{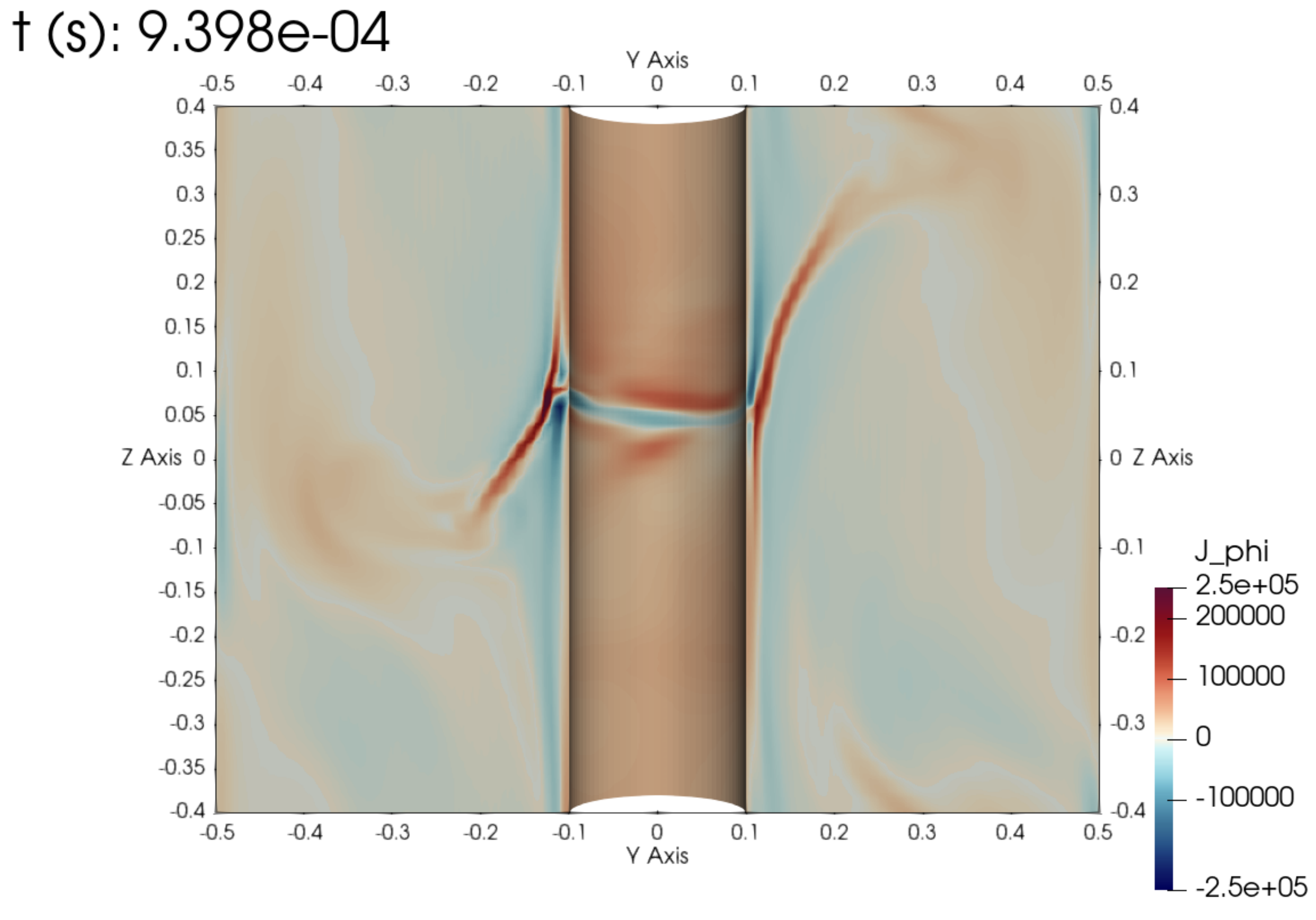}{.45\textwidth}{(a)}
    \fig{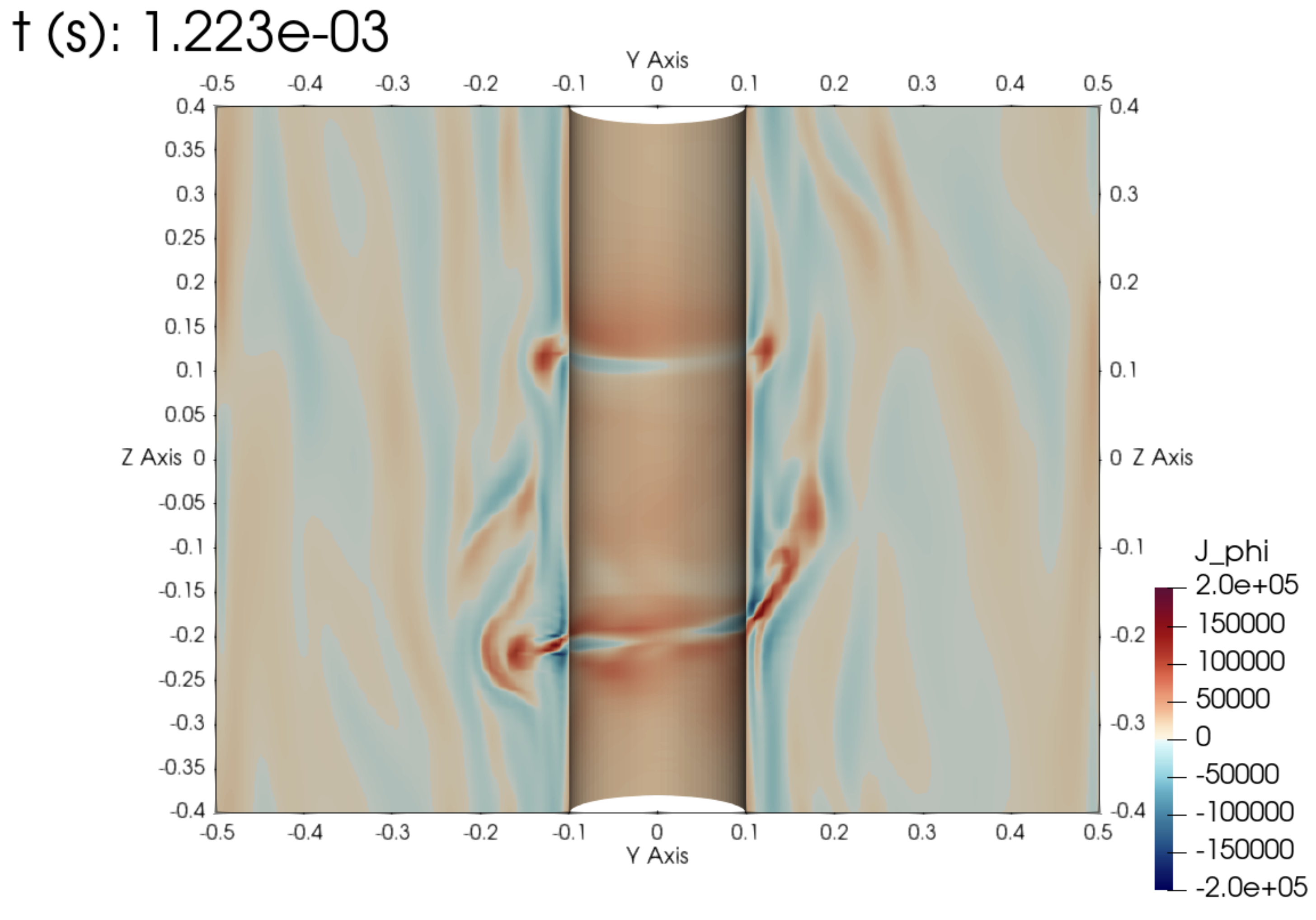}{.45\textwidth}{(d)}}
    
    \gridline{\fig{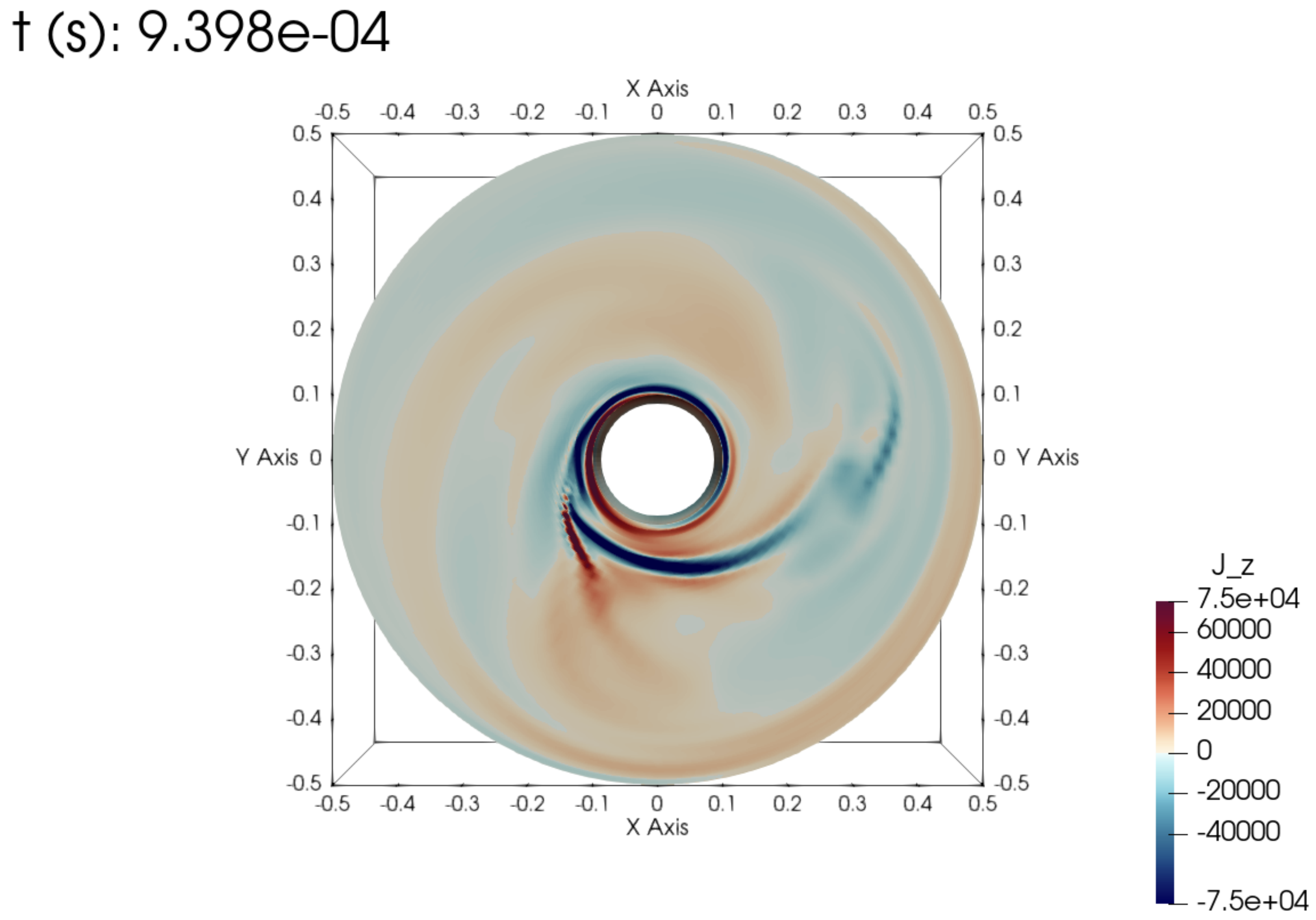}{.45\textwidth}{(b)}
    \fig{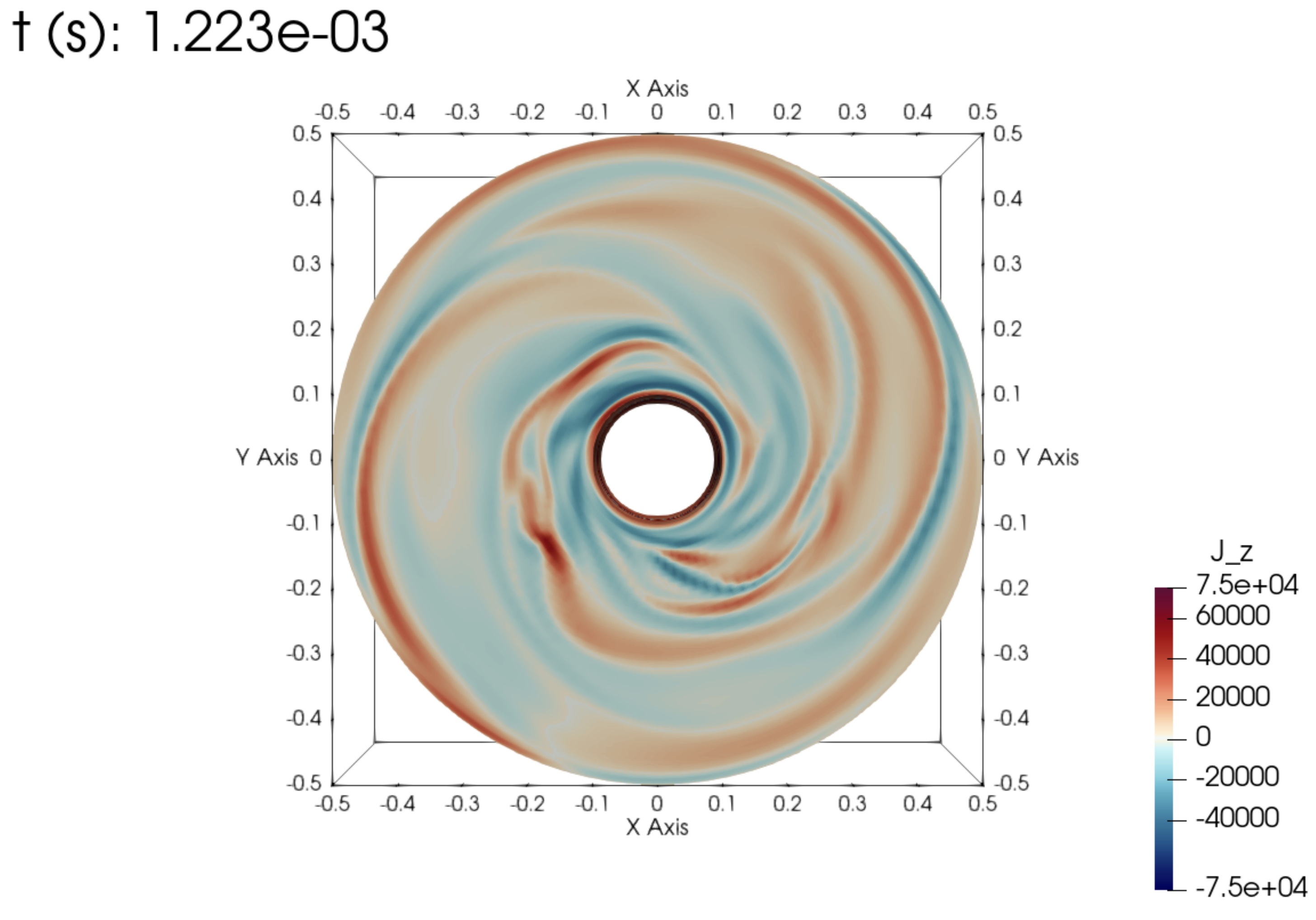}{.45\textwidth}{(e)}}
    
    \gridline{\fig{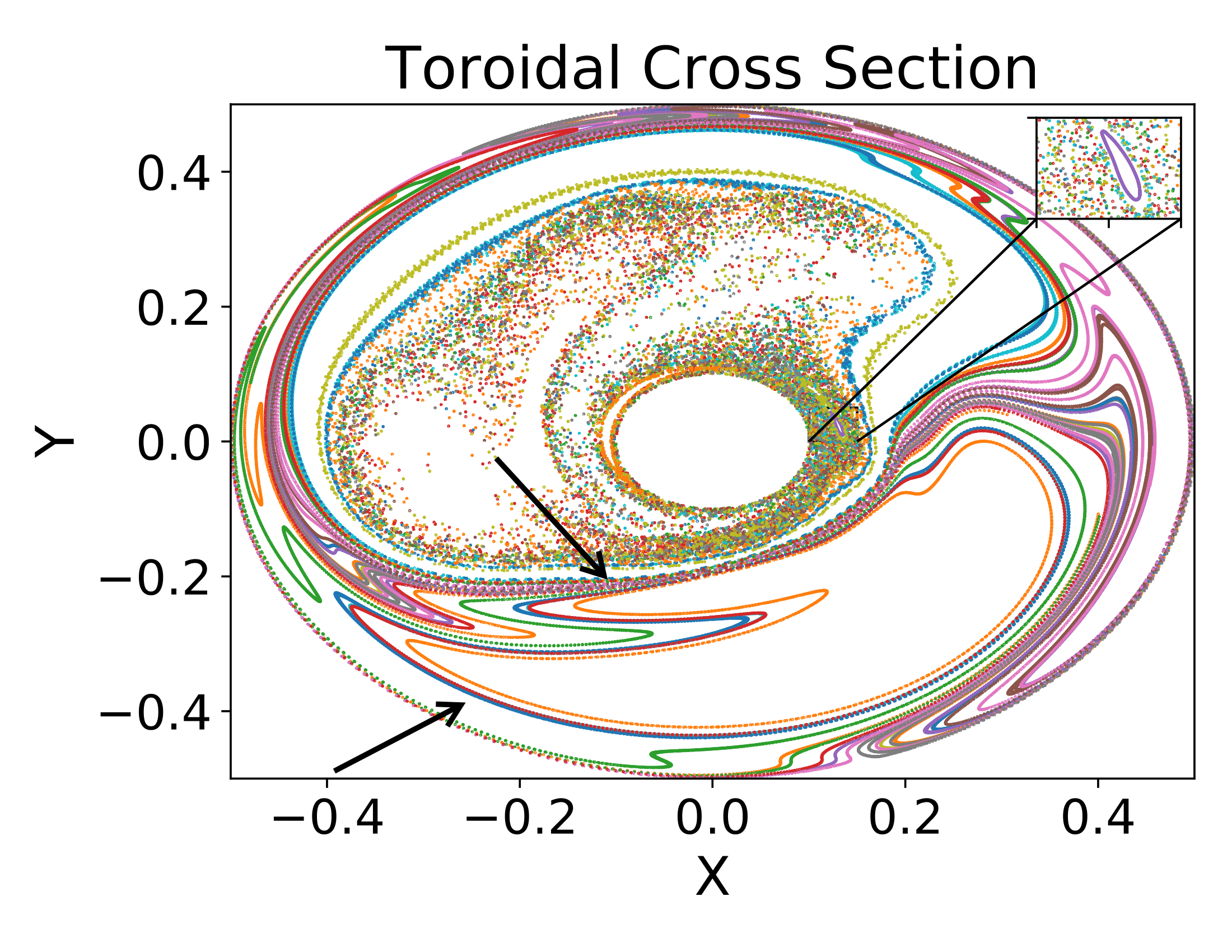}{.45\textwidth}{(c)}
    \fig{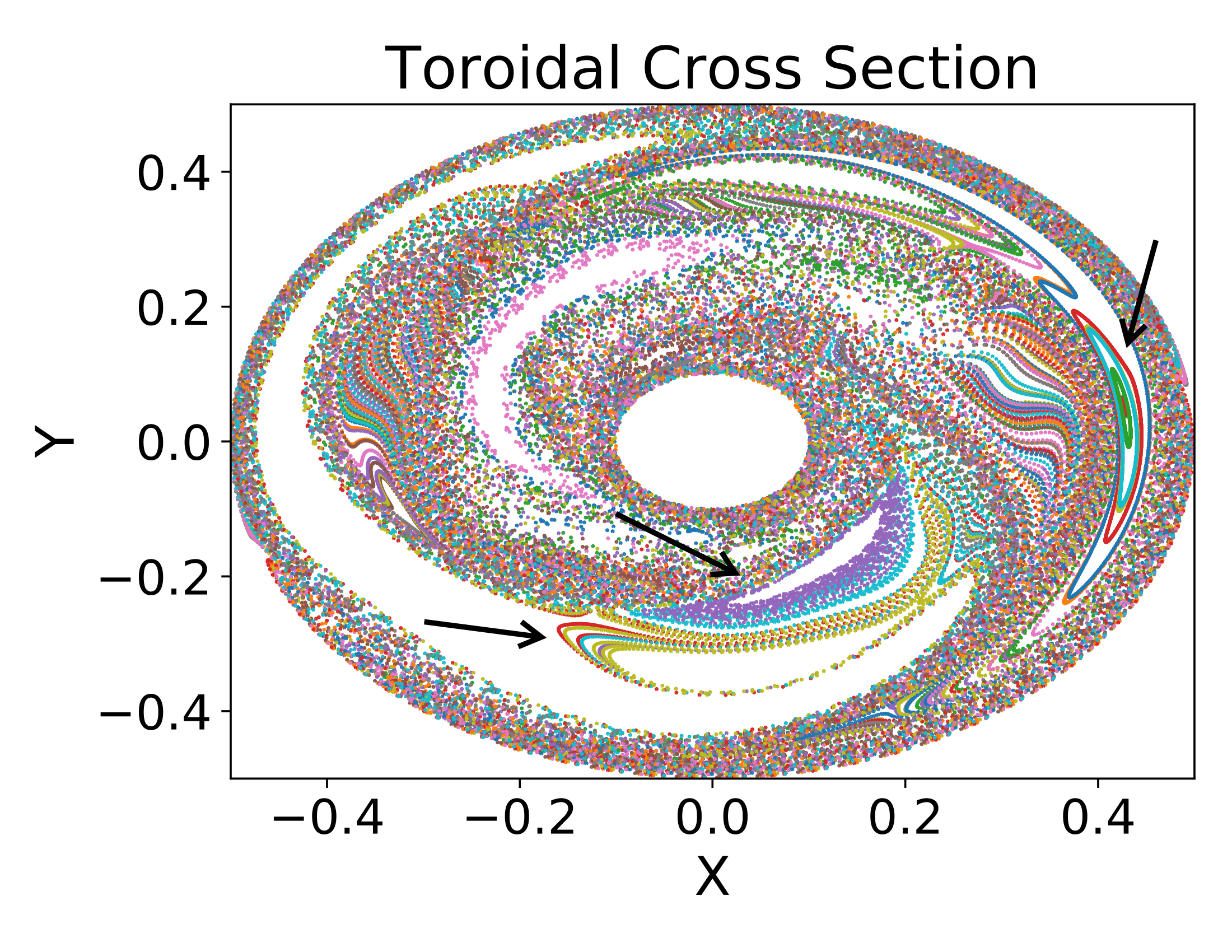}{.45\textwidth}{(f)}}
    
    \caption{{Current densities and Poincare plots for the 3D case during saturation at two times, left column, (a)-(c), are taken at $t=\SI{0.9398}{ms}$, and in the right, (d)-(f), are taken at $t=\SI{1.223}{ms}$. (a) and (d) are the vertically extended current densities $J_\phi$ in a poloidal cut. (b) and (e) are the vertical current densities $J_z$ in a toroidal cut. (c) and (f) are the Poincar\'e plots in the toroidal plane. The black arrows points to the large magnetic island structure.}}
    
    \label{fig:3d_turb}
\end{figure*}

As the higher non-axisymmetric $n$ toroidal modes grow and saturate, the plasma flow becomes more turbulent. The remaining current sheets in the poloidal plane at this time have merged and now get stretched vertically (see figure \ref{fig:3d_turb}a), becoming a 3D current sheet, or filament. These filaments are later dissipated. Interestingly, though we do not show it here, Poincar\'e plots of the magnetic field in the poloidal plane reveal that a large island forms around this time near the radial edge of the disk.

Plasma current activity also rises in the toroidal plane, as seen in figure \ref{fig:3d_turb}b, with toroidally extending $J_z$ sheets swirling around the center of the disk and diffusing outward. In time this swirl of $J_z$ rotates around the center of the disk. This is consistent with the work of \citet{Machida_2003}, who also observed spiraling current density in the in-fall region of the disk. The appearance of this current density $J_z$ is significant, as current in the $z$ direction is important for current outflows in astrophysical jets {\citep[and references therein]{dal2005astrophysical, blackman2020persistent}}.

The Poincar\'e plot in the toroidal plane at this time, seen in figure \ref{fig:3d_turb}c also exhibits magnetic reconnection activities. First, there are small islands near the center of the disk. This is due to the reconnection occurring along the vertical current sheets in the inner region seen in figure \ref{fig:3d_turb}b. Second, a large-scale closed flux surface also appears, which 
is due to the $n=1$ non-axisymmetric mode activity. This structure over time compresses into a smaller, double-island structure as turbulence evolves. 

Towards the end of the simulation, as the higher modes become dominant, particularly the $n=2$ and $n=3$ modes, toroidal current sheets reappear in the poloidal plane, as seen in figure \ref{fig:3d_turb}d at $t=\SI{1.22}{ms}$. These sheets are far more dynamic at this time, however, due to the turbulent flows. They flap up and down, and every so often they eject current in a cloud (currently happening in the bottom left sheet in figure \ref{fig:3d_turb}d). At this time, additional turbulent spirals of $J_z$ also appear near the center in the toroidal plane as well, as seen in figure \ref{fig:3d_turb}e. More large islands spawn in the toroidal plane, as seen in figure \ref{fig:3d_turb}f. These 3D island structures are {likely} generated due to a combination of current sheet reconnection and turbulent driven magnetic reconnection {\citep{Lazarian_1999}} from $n=2$ and $n=3$ fluctuations.

\section{Summary and Conclusions}

In the framework of a resistive MHD model, we have numerically studied the evolution of current and magnetic field in 2D accretion disks and a 3D disk subject to a vertical seed magnetic field. In doing so we have made four key observations.
First, we found that MRI induces radial perturbations in the magnetic field, creating toroidal Harris type current sheets. This proves that MRI is a driver for current sheets in astrophysical disks (as we call MRI-driven current sheets).
Second, these current sheets are drawn to one another by a Lorentz force and merge, driving magnetic reconnection and local momentum transport at $S > \num{3e2}$. This process is a source of plasmoids that could be contributing to flares and gamma-ray emissions {\citep{Singh_2015,  Ripperda_2020}} and extreme particle acceleration {\citep{uzdensky2011reconnection, cerutti2014gamma, Sironi2016}}.

Third, at $S \geq \num{5e3}$, the current sheets become thin and subject to plasmoid instability driven by MRI, breaking into multiple plasmoids. This is significant, as it shows that MRI itself can lead to plasmoid instability, even in the absence of other effects such as general relativity or non-ideal MHD. 
{This is the first demonstration of formation of current sheets driven by large-scale primary (axisymmetric) MRI mode (see figure \ref{fig:elecd5_pan}a), and the subsequent breaking of these current sheets due to spontaneous plasmoid reconnection (tearing instability \citep{FKR_1963} in a current sheet).} {Our calculation of reconnection times also approximately follows the theoretical scaling of $S^{1/2}$ at low S, but starts to show weak dependency on S with the onset of plasmoid instability.} The impact of the plasmoid instability on MRI saturation mechanism and its effect on  momentum transport will remain for a future study. 

Finally, once non-axisymmetric effects become significant, turbulence stretches the remaining current sheets vertically and dissipates them, and large magnetic islands form in the outer regions of the disk in both the poloidal and toroidal plane. In addition, smaller plasmoids form near the center of the disk in the toroidal plane due to reconnection along vertical current sheets. These vertical currents (generated by the large-scale azimuthal dynamo fields) are important for current outflows in astrophysical jets and the disk corona region {\citep{dal2005astrophysical, mckinney2009stability, blackman2020persistent}}. Eventually, toroidal current sheets reappear in the poloidal plane but are unstable to the plasma flow. 

In summary, {based on a first-principle approach using resistive MHD solutions, it is found that initial fields lines threading a differentially rotating disk are stretched radially and toroidally outward by the primary MRI mode\footnote{for visualizations of the field lines see Fig. 5 in ref.~\citet{ebrahimi_blackman_2016}} to produce radially extended azimuthal current sheets (and reconnection sites with oppositely directed field lines), which break into plasmoids at high local Lundquist number.} With our cylindrical disk model, we have uncovered some of the fundamental reconnection processes triggered or driven by MRI. By examining the current evolution in the poloidal and toroidal planes and following the magnetic field lines in both 2D and 3D simulations, we have identified sites of fast plasmoid reconnection 
{\citep{Bhatteracharjee2009,Huang_Bhattacharjee_2010,Loureiro_2012,ebrahimi_2015}, as well as 3D reconnection sites potentially due to turbulent reconnection  and nonaxisymmetric fluctuations.
In this paper, we haven't focused on small-scale MRI turbulence and the associated power spectrum.
Although turbulent reconnection \citep{Lazarian_1999, Kowal_2009, Eyink_2011, kadowaki_2018, Lazarian_2020, Jafari_2020} is relevant, it is not the primary driver of reconnection in our 2D simulations. In our 3D simulations, however, small-scale turbulence could play a role. For example, small-scale 3D disturbances around the Harris type current sheets could enhance local $S$ to trigger fast reconnection \citep{Ebrahimi_pop_2016}. Here on the other hand, it is hard to differentiate between the driven turbulence reconnection vs the spontaneous plasmoid reconnection due to MRI. We believe both occur in our 3D simulations, but further simulations are required to identify the role of small-scale turbulence on the reconnection rate.}

Lastly, we should note that our cylindrical disk model may have some limitations, but on the other hand provides solutions in a real global domain with curvature terms and explicit resistivity and viscosity. Our focus here has been to capture the associated reconnection physics of MRI in cylindrical differentially rotating plasmas. Further detailed 3D simulations including other physics such as non-ideal effects, stratification, and higher magnetic Reynolds number will be investigated in future work.

\begin{acknowledgements}
This work was supported in part by the U.S. Department of Energy, Office of Science, Office of Workforce Development for Teachers and Scientists (WDTS) under the Science Undergraduate Laboratory Internships Program (SULI). 
FE also acknowledges the Max-Planck/Princeton Center for Plasma Physics. Computations were performed at NERSC and local PPPL cluster.

We would also like to thank our anonymous referee for providing useful feedback.
\end{acknowledgements}

\bibliography{references}
\bibliographystyle{aasjournal}

\end{document}